\documentclass{aastex61}
\pdfoutput=1
\usepackage{amsmath,amstext}
\usepackage[T1]{fontenc}
\usepackage{apjfonts} 
\usepackage{graphicx}
\usepackage{rotating}

\shorttitle{AASTeX 6.1 Template}
\shortauthors{Yang et al.}

\begin{document}

\title{High-Resolution Near-Infrared Polarimetry and Sub-Millimeter Imaging of FS Tau A: Possible Streamers in Misaligned Circumbinary Disk System}

\correspondingauthor{Yi YANG}
\email{yi.yang@nao.ac.jp}
\author{Yi YANG}
\affiliation{Astrobiology Center, NINS, 2-21-1, Osawa, Mitaka, Tokyo 181-8588, Japan}
\affiliation{National Astronomical Observatory of Japan (NAOJ), National Institutes of Natural Sciences (NINS), 2-21-1 Osawa, Mitaka, Tokyo 181-8588,Japan}

\author{Eiji AKIYAMA}
\affiliation{Institute for the Advancement of Higher Education, Hokkaido University, Kita 17, Nishi 8, Kita-ku, Sapporo, Hokkaido, 060-0817, Japan}

\author{Thayne CURRIE}
\affiliation{Subaru Telescope, NAOJ, NINS, 650 North A'ohoku Place, Hilo, HI 96720, USA}

\author{Ruobing DONG}
\affiliation{Department of Physics \& Astronomy, University of Victoria, Victoria BC V8P 1A1, Canada}

\author{Jun HASHIMOTO}
\affiliation{Astrobiology Center, NINS, 2-21-1, Osawa, Mitaka, Tokyo 181-8588, Japan}
\affiliation{National Astronomical Observatory of Japan (NAOJ), National Institutes of Natural Sciences (NINS), 2-21-1 Osawa, Mitaka, Tokyo 181-8588,Japan}

\author{Saeko S. HAYASHI}
\affiliation{Department of Astronomical Science, SOKENDAI (The Graduate University for Advanced Studies), 2-21-1 Osawa, Mitaka, Tokyo 181-8588, Japan}
\affiliation{National Astronomical Observatory of Japan (NAOJ), National Institutes of Natural Sciences (NINS), 2-21-1 Osawa, Mitaka, Tokyo 181-8588,Japan}

\author{Carol A. GRADY}
\affiliation{Exoplanets and Stellar Astrophysics Laboratory, Code 667, Goddard Space Flight Center, Greenbelt, MD, 20771, USA}
\affiliation{Eureka Scientific, 2452 Delmer, Suite 100, Oakland CA 96002, USA}
\affiliation{Goddard Center for Astrobiology, NASA Goddard Space Flight Center, Greenbelt, MD 20771, USA}

\author{Markus JANSON}
\affiliation{Department of Astronomy, Stockholm University, AlbaNova University Center, SE-106 91 Stockholm, Sweden}

\author{Nemanja JOVANOVIC}
\affiliation{Jet Propulsion Laboratory, California Institute of Technology, M/S 171-113 4800 Oak Grove Drive Pasadena, CA 91109 USA}

\author{Taichi UYAMA}
\affiliation{Department of Astronomy, The University of Tokyo, 7-3-1 Hongo, Bunkyo-ku, Tokyo 113-0033, Japan}

\author{Takao NAKAGAWA}
\affiliation{Institute of Space and Astronautical Science, Japan Aerospace Exploration Agency, 3-1-1 Yoshinodai, Chuo-ku, Sagamihara, Kanagawa 252-5210, Japan}

\author{Tomoyuki KUDO}
\affiliation{Astrobiology Center, NINS, 2-21-1, Osawa, Mitaka, Tokyo 181-8588, Japan}
\affiliation{National Astronomical Observatory of Japan (NAOJ), National Institutes of Natural Sciences (NINS), 2-21-1 Osawa, Mitaka, Tokyo 181-8588,Japan}

\author{Nobuhiko KUSAKABE}
\affiliation{Astrobiology Center, NINS, 2-21-1, Osawa, Mitaka, Tokyo 181-8588, Japan}
\affiliation{National Astronomical Observatory of Japan (NAOJ), National Institutes of Natural Sciences (NINS), 2-21-1 Osawa, Mitaka, Tokyo 181-8588,Japan}

\author{Masayuki KUZUHARA}
\affiliation{Astrobiology Center, NINS, 2-21-1, Osawa, Mitaka, Tokyo 181-8588, Japan}
\affiliation{National Astronomical Observatory of Japan (NAOJ), National Institutes of Natural Sciences (NINS), 2-21-1 Osawa, Mitaka, Tokyo 181-8588,Japan}

\author{Lyu ABE}
\affiliation{Laboratoire Lagrange (UMR 7293), Universite de Nice-Sophia Antipolis, CNRS, Observatoire de la Coted'azur, 28 avenue Valrose, 06108 Nice Cedex 2, France}

\author{Wolfgang BRANDNER}
\affiliation{Max Planck Institute for Astronomy, K{\"o}nigstuhl 17, 69117 Heidelberg, Germany}

\author{Timothy D. BRANDT}
\affiliation{Department of Physics, Broida Hall, University of California, Santa Barbara, CA 93106-9530}

\author{Michael BONNEFOY}
\affiliation{Univ. Grenoble Alpes, CNRS, IPAG, F-38000 Grenoble, France}

\author{Joseph C. CARSON}
\affiliation{Department of Physics and Astronomy, College of Charleston, 58 Coming St., Charleston, SC 29424, USA}

\author{Jeffrey CHILCOTE}
\affiliation{Department of Physics, University of Notre Dame, 225 Nieuwland Science Hall, Notre Dame, IN 46556, USA}

\author{Evan A. RICH}
\affiliation{H. L. Dodge Department of Physics \& Astronomy, University of Oklahoma, 440 W Brooks St., Norman, OK 73019, USA}

\author{Markus FELDT}
\affiliation{Max Planck Institute for Astronomy, K{\"o}nigstuhl 17, 69117 Heidelberg, Germany}

\author{Miwa GOTO}
\affiliation{Universit{\"a}ts-Sternwarte M{\"u}nchen, Ludwig-Maximilians-Universit{\"a}t, Scheinerstr. 1, 81679 M{\"u}nchen, Germany}

\author{Tyler GROFF}
\affiliation{NASA-Goddard Space Flight Center, Greenbelt, MD, USA}

\author{Olivier GUYON}
\affiliation{Astrobiology Center, NINS, 2-21-1, Osawa, Mitaka, Tokyo 181-8588, Japan}
\affiliation{Subaru Telescope, NAOJ, NINS, 650 North A'ohoku Place, Hilo, HI 96720, USA}
\affiliation{Steward Observatory, University of Arizona, 933 N Cherry Ave., Tucson AZ 85719, USA}

\author{Yutaka HAYANO}
\affiliation{Department of Astronomical Science, SOKENDAI (The Graduate University for Advanced Studies), 2-21-1 Osawa, Mitaka, Tokyo 181-8588, Japan}
\affiliation{National Astronomical Observatory of Japan (NAOJ), National Institutes of Natural Sciences (NINS), 2-21-1 Osawa, Mitaka, Tokyo 181-8588,Japan}
\affiliation{Subaru Telescope, NAOJ, NINS, 650 North A'ohoku Place, Hilo, HI 96720, USA}

\author{Masahiko HAYASHI}
\affiliation{JSPS Bonn Office, Wissenschaftszentrum, Ahrstrasse 58, 53175 Bonn, Germany}

\author{Thomas HENNING}
\affiliation{Max Planck Institute for Astronomy, K{\"o}nigstuhl 17, 69117 Heidelberg, Germany}

\author{Klaus W. HODAPP}
\affiliation{Institute for Astronomy, University of Hawaii, 640 N. A'ohoku Place, Hilo, HI 96720, USA}

\author{Miki ISHII}
\affiliation{National Astronomical Observatory of Japan (NAOJ), National Institutes of Natural Sciences (NINS), 2-21-1 Osawa, Mitaka, Tokyo 181-8588,Japan}

\author{Masanori IYE}
\affiliation{National Astronomical Observatory of Japan (NAOJ), National Institutes of Natural Sciences (NINS), 2-21-1 Osawa, Mitaka, Tokyo 181-8588,Japan}

\author{Ryo KANDORI}
\affiliation{Astrobiology Center, NINS, 2-21-1, Osawa, Mitaka, Tokyo 181-8588, Japan}

\author{Jeremy KASDIN}

\author{Gillian R. KNAPP}
\affiliation{Department of Astrophysical Science, Princeton University, Peyton Hall, Ivy Lane, Princeton, NJ 08544, USA}

\author{Jungmi KWON}
\affiliation{Department of Astronomy, The University of Tokyo, 7-3-1 Hongo, Bunkyo-ku, Tokyo 113-0033, Japan}

\author{Julien LOZI}
\affiliation{Subaru Telescope, NAOJ, NINS, 650 North A'ohoku Place, Hilo, HI 96720, USA}

\author{Frantz MARTINACHE}
\affiliation{ Universit\'e C\^ote d'Azur, Observatoire de la C\^ote d'Azur, CNRS, Laboratoire Lagrange, France}

\author{Taro MATSUO}
\affiliation{Department of Earth and Space Science, Graduate School of Science, Osaka University, 1-1 Machikaneyamacho, Toyonaka, Osaka 560-0043, Japan}

\author{Satoshi MAYAMA}
\affiliation{SOKENDAI(The Graduate University for Advanced Studies), Shonan International Village, Hayama-cho, Miura-gun, Kanagawa 240-0193, Japan}

\author{Michael W. MCELWAIN}
\affiliation{Exoplanets and Stellar Astrophysics Laboratory, Code 667, Goddard Space Flight Center, Greenbelt, MD, 20771, USA}

\author{Shoken MIYAMA}
\affiliation{Hiroshima University, 1-3-2 Kagamiyama, Higashihiroshima, Hiroshima 739-8511, Japan}

\author{Jun-Ichi MORINO}
\affiliation{National Astronomical Observatory of Japan (NAOJ), National Institutes of Natural Sciences (NINS), 2-21-1 Osawa, Mitaka, Tokyo 181-8588,Japan}

\author{Amaya MORO-MARTIN}
\affiliation{Space Telescope Science Institute(STScI), 3700 San Martin Drive, Baltimore, MD 21218}

\author{Tetsuo NISHIMURA}
\affiliation{Subaru Telescope, NAOJ, NINS, 650 North A'ohoku Place, Hilo, HI 96720, USA}

\author{Tae-Soo PYO}
\affiliation{Department of Astronomical Science, SOKENDAI (The Graduate University for Advanced Studies), 2-21-1 Osawa, Mitaka, Tokyo 181-8588, Japan}
\affiliation{Subaru Telescope, NAOJ, NINS, 650 North A'ohoku Place, Hilo, HI 96720, USA}

\author{Eugene SERABYN}
\affiliation{Jet Propulsion Laboratory, California Institute of Technology, M/S 183-900 4800 Oak Grove Drive Pasadena, CA 91109, USA}

\author{Hiroshi SUTO}
\affiliation{Astrobiology Center, NINS, 2-21-1, Osawa, Mitaka, Tokyo 181-8588, Japan}
\affiliation{National Astronomical Observatory of Japan (NAOJ), National Institutes of Natural Sciences (NINS), 2-21-1 Osawa, Mitaka, Tokyo 181-8588,Japan}

\author{Ryuji SUZUKI}
\affiliation{National Astronomical Observatory of Japan (NAOJ), National Institutes of Natural Sciences (NINS), 2-21-1 Osawa, Mitaka, Tokyo 181-8588,Japan}

\author{Michihiro TAKAMI}
\affiliation{Institute of Astronomy and Astrophysics, Academia Sinica, P.O. Box 23-141, Taipei 10617, Taiwan}

\author{Naruhisa TAKATO}
\affiliation{Department of Astronomical Science, SOKENDAI (The Graduate University for Advanced Studies), 2-21-1 Osawa, Mitaka, Tokyo 181-8588, Japan}
\affiliation{Subaru Telescope, NAOJ, NINS, 650 North A'ohoku Place, Hilo, HI 96720, USA}

\author{Hiroshi TERADA}
\affiliation{National Astronomical Observatory of Japan (NAOJ), National Institutes of Natural Sciences (NINS), 2-21-1 Osawa, Mitaka, Tokyo 181-8588,Japan}

\author{Christian THALMANN}
\affiliation{Swiss Federal Institute of Technology (ETH Zurich), Institute for Astronomy, Wolfgang-Pauli-Strasse 27, CH-8093 Zurich, Switzerland}

\author{Edwin L. TURNER}
\affiliation{Department of Astrophysical Science, Princeton University, Peyton Hall, Ivy Lane, Princeton, NJ 08544, USA}
\affiliation{Kavli Institute for Physics and Mathematics of the Universe, The University of Tokyo, 5-1-5 Kashiwanoha, Kashiwa, Chiba 277-8568, Japan}

\author{Makoto WATANABE}
\affiliation{Department of Cosmosciences, Hokkaido University, Kita-ku, Sapporo, Hokkaido 060-0810, Japan}

\author{John P. WISNIEWSKI}
\affiliation{H. L. Dodge Department of Physics \& Astronomy, University of Oklahoma, 440 W Brooks St., Norman, OK 73019, USA}

\author{Toru YAMADA}
\affiliation{Astronomical Institute, Tohoku University, Aoba-ku, Sendai, Miyagi 980-8578, Japan}

\author{Hideki TAKAMI}
\affiliation{Department of Astronomical Science, SOKENDAI (The Graduate University for Advanced Studies), 2-21-1 Osawa, Mitaka, Tokyo 181-8588, Japan}
\affiliation{National Astronomical Observatory of Japan (NAOJ), National Institutes of Natural Sciences (NINS), 2-21-1 Osawa, Mitaka, Tokyo 181-8588,Japan}

\author{Tomonori USUDA}
\affiliation{Department of Astronomical Science, SOKENDAI (The Graduate University for Advanced Studies), 2-21-1 Osawa, Mitaka, Tokyo 181-8588, Japan}
\affiliation{National Astronomical Observatory of Japan (NAOJ), National Institutes of Natural Sciences (NINS), 2-21-1 Osawa, Mitaka, Tokyo 181-8588,Japan}

\author{Motohide TAMURA}
\affiliation{Department of Astronomy, The University of Tokyo, 7-3-1 Hongo, Bunkyo-ku, Tokyo 113-0033, Japan}
\affiliation{Astrobiology Center, NINS, 2-21-1, Osawa, Mitaka, Tokyo 181-8588, Japan}
\affiliation{National Astronomical Observatory of Japan (NAOJ), National Institutes of Natural Sciences (NINS), 2-21-1 Osawa, Mitaka, Tokyo 181-8588,Japan}

\begin{abstract}

We analyzed the young (2.8-Myr-old) binary system FS Tau A using near-infrared (H-band) high-contrast polarimetry data from Subaru/HiCIAO and sub-millimeter CO (J=2-1) line emission data from ALMA. Both the near-infrared and sub-millimeter observations reveal several clear structures extending to $\sim$240 AU from the stars. Based on these observations at different wavelengths, we report the following discoveries. One arm-like structure detected in the near-infrared band initially extends from the south of the binary with a subsequent turn to the northeast, corresponding to two bar-like structures detected in ALMA observations with an LSRK velocity of 1.19-5.64 km/s. Another feature detected in the near-infrared band extends initially from the north of the binary, relating to an arm-like structure detected in ALMA observations with an LSRK velocity of 8.17-16.43 km/s. From their shapes and velocities, we suggest that these structures can mostly be explained by two streamers that connect the outer circumbinary disk and the central binary components. These discoveries will be helpful for understanding the evolution of streamers and circumstellar disks in young binary systems.

\end{abstract}


\section{Introduction}
Over 150 planets (e.g., $\gamma$ Cep Ab, \citet{Hatzes2003}; Kepler 16 b, \citet{Doyle2011}; ROXs 42Bb, \citet{Currie2014}) in binary or multiple systems have been confirmed\footnote{Catalogue of Exoplanets in Binary Star Systems: https://www.univie.ac.at/adg/schwarz/multiple.html, \citep{Schwarz2016}.}. To understand their formation, it is important to understand binary-disk interactions by investigating protoplanetary disks in binary systems. Theoretical works have shown that binaries have various effects on disks, such as inducing misalignment in disks with the binary orbital plane (e.g., \citet{Martin2017}), opening gaps in circumbinary disks (e.g., \citet{Artymowicz1994}), and driving spiral arms (e.g., \citet{Dong2016}). A binary can trigger streamers inside the opened gaps in disks, bringing materials to the region near the binary \citep[e.g., ][]{Nelson2016, Yang2017}. This is believed to help sustain the circumstellar disks around stars, facilitating planet formation. Therefore, research on streamers in binary systems is useful for understanding the formation of S-type planets, i.e., planets that orbit one star in a binary system.

FS Tau, also known as Haro 6-5, is a young multiple T-Tauri star system with an age of about 2.8 Myr \citep{Palla2002a}. No Gaia distance is available for this target. Therefore, the typical distance for the Taurus star formation region \footnote{140 pc, \citep[e.g.,][]{Kenyon1994}} is used in this work. FS Tau consists of FS Tau A and FS Tau B. FS Tau A, itself a binary system, has a total mass of approximately 0.78$\pm0.25M_\odot$, a semi-major axis of about 0.$\arcsec$275, and an eccentricity of 0.168 \citep{Tamazian2002}. FS Tau B is a single star located at about 20$\arcsec$ west of FS Tau A; it is famous for its bipolar outflows, which have been extensively studied\citep[e.g.,][]{Liu2012}. However, research on FS Tau A is limited. FS Tau A has an accretion rate (log($\dot{M}/M_\odot yr^{-1}$)$\sim$-9.5, \citet{White2001}), making it a potential target for finding streamers. \citet{Krist1998} did not find clear evidence of a disk around this binary from their Hubble Space Telescope Wide Field and Planetary Camera 2 (WFPC2) observations. \citet{Hioki2011} suggested a large circumbinary disk extending to about 630 AU and inclined by 30$^\circ$ to 40$^\circ$ based on Subaru Telescope/Coronagraphic Imager with Adaptive Optics (CIAO) H-band observations as well as Hubble Space Telescope/Advanced Camera for Surveys 606W-band polarimetric images. Its southeast side is likely to be closer to us because it is brighter than the northwest side in the H-band due to the forward scattering of dust. However, \citet{Hioki2011} only resolved the structures beyond 0.$\arcsec$8 from the stars. To understand how a young binary interacts with its surrounding disks, it is necessary to observe such systems at smaller inner working angles. By using the high-contrast instrument (HiCIAO) and the adaptive optics system of the Subaru Telescope (AO188), we will be able to resolve structures down to 0.$\arcsec$1 from the star, and ALMA observations can provide image of gas at the similar spatial resolution level as the near-infrared observations to help understand the dynamics in the disk. Therefore, using HiCIAO and ALMA observations, we may be able to resolve the inner region of the circumbinary disk for the first time.  

In this paper, we present and discuss the Subaru/HiCIAO near-infrared polarimetric data and Atacama Large Millimeter/submillimeter Array (ALMA) CO ($J$=2-1) spectral line data of FS Tau A, both of which resolved structures near FS Tau A. The rest of this paper is organized as follows. Section 2 introduces the observations and data reduction. Section 3 shows the results. Section 4 discusses the disk and streamer structures. Section 5 gives the conclusions.

\section{Observations and Data Reduction}
The H-band near-infrared observations of FS Tau A were taken on Dec. 26, 2011, using the HiCIAO near-infrared camera and the adaptive optics instrument AO188 \citep{Hayano2010} mounted on the Subaru Telescope.
This observation was part of the survey program Strategic Explorations of Exoplanets
and Disks with Subaru (SEEDS), which began in 2009. The ``natural" seeing during the observation was about 0.$\arcsec$9, while the AO corrected PSF has a FWHM of about 0.$\arcsec$3.
This observation employed the standard polarized differential imaging mode, which
uses a Wollaston prism to split the light into two $2048\times1024$-pixel
channels on the detector, corresponding to o- and e-polarizations, respectively. A half-wave
plate was used in the observation. It was rotated to position
angles of $0^{\circ}$, $22.5^{\circ}$, $45^{\circ}$, and $67.5^{\circ}$
to measure the Stokes parameters. This cycle was repeated 13 times during the observation. A total of 52 frames were collected, each with
an exposure time of 50 s, for a total integration time of about 43 minutes. 


As for the data reduction of FS Tau A, firstly the stripes were removed and flat field and bad pixels were corrected. The pixel scale and detector orientation were derived by taking the image of the global cluster M15 which was taken within a few days in the same run, and comparing it with the archival M15 image previously taken by Advanced Camera for Surveys (ACS) mounted on Hubble Space Telescope \citep{vandermarel2007}, assuming the latter was distortion free. A few hundred stars were used for the calibration. The pixel scale of HiCIAO detector has $\sim$3\% difference between the horizontal and vertical axes, and there is $\sim0.3^\circ$ offset between the vertical axis and the celestial north. Then we used bilinear interpolation to correct these distortions via the Image Reduction and Analysis Facility (IRAF) \textit{geomap} and \textit{geotran} commands. The pixel scale was corrected to 9.500$\pm$0.005 mas/pixel and the detector’s offset to the celestial north was corrected to 0$\pm$0.02$^\circ$. The accuracy of the image registration is less than 4 mas within the central 10$\arcsec\times$10$\arcsec$ of the FoV.
After these steps, the images were first cross-correlated in different channels then we derived the Stokes parameters $+Q$, $+U$, $-Q$, and $-U$ by subtracting the e-images from the o-images. In the next step, the $Q$ and $U$ images were constructed as $Q=((+Q)-(-Q))/2$, $U=((+U)-(-U))/2$. The instrumental polarization was corrected based on the method described by \citet{Joos2008}. The observed Stokes $Q/U$ parameters can be described as a linear combination of true Stokes $I/Q/U$ parameters: 
\begin{equation}
\begin{aligned}
Q_{Observed} &= a*I + b*Q + c*U\\
U_{Observed} &= d*I + e*Q + f*U
\end{aligned}
\end{equation}
here a-f are coefficients calibrated during observations for instrumental polarization corrections. The true Stokes parameters can then be derived by solving such a combination of equations. The Stokes $I$ image, or intensity image, was derived by averaging the sum of the o- and e- images in all frames. 

The ALMA archival data at Band 6 (211-275 GHz) toward FS Tau A (Project ID 2013.1.00105.S) is used in the paper. It was observed on Sep. 19, 2015 with 36 antennas whose baselines range from 41.4 m to 2.3 km, resulting in a maximum recoverable size of 3.$\arcsec$88. The 2SB receivers are tuned at the rest frequency of 230.538 GHz for CO ($J$=2-1) emissions with the bandwidth of 23.92 MHz and the spectral resolution is 976.54 kHz, corresponding to 1264.70m/s in the velocity resolution. The integration time on source after data flagging is 120.960 s and the obtained beam size is 0.$\arcsec16\times0.\arcsec$22. The average precipitable water vapor (PWV) during the observation was about 1.74 mm. The calibration was done by the pipeline in the Common Astronomy Software Applications (CASA) package 4.3.1 and imaging was done by CLEAN task provided in CASA. J0510+1800 is used for bandpass calibration, J0510+180 is used for flux calibration and J0426+2327 is used for phase calibration. Note that the continuum image of FS Tau A is not strong enough to make self-calibration safely so we do not apply it in CO ($J$=2-1) line imaging.

\section{Results}

\subsection{Near-Infrared Observations}

\subsubsection{Binary Orbit}

\begin{figure}[ht!]
	\figurenum{1}
	\gridline{
    \fig{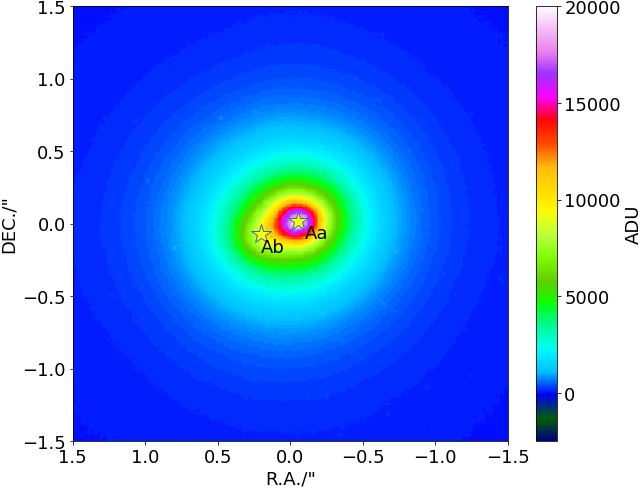}{0.45\textwidth}{a}
    \label{fig:fstaui}
    \fig{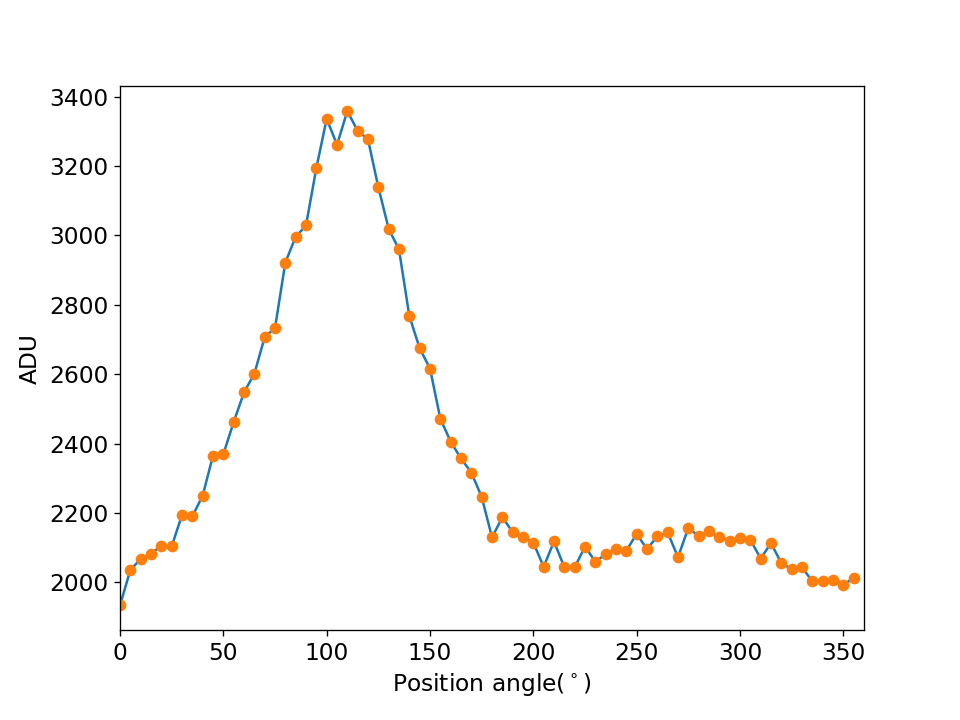}{0.45\textwidth}{b}
    \label{fig:azipro}
	}
    \gridline{
    \fig{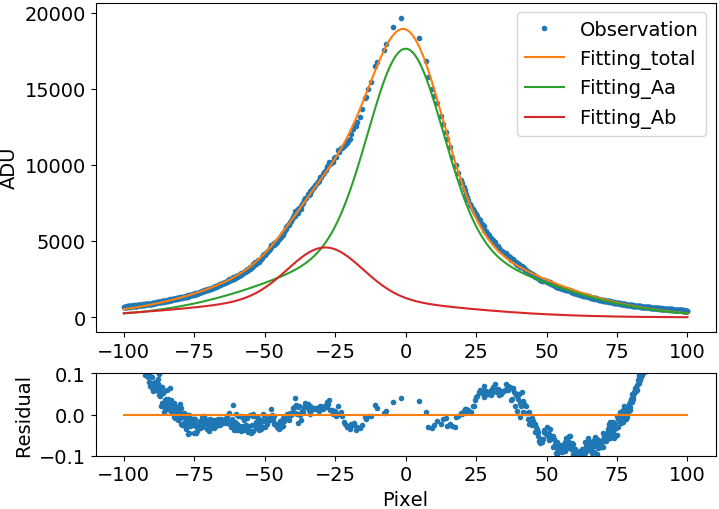}{0.4\textwidth}{c}
    \label{fig:radpro}
    \fig{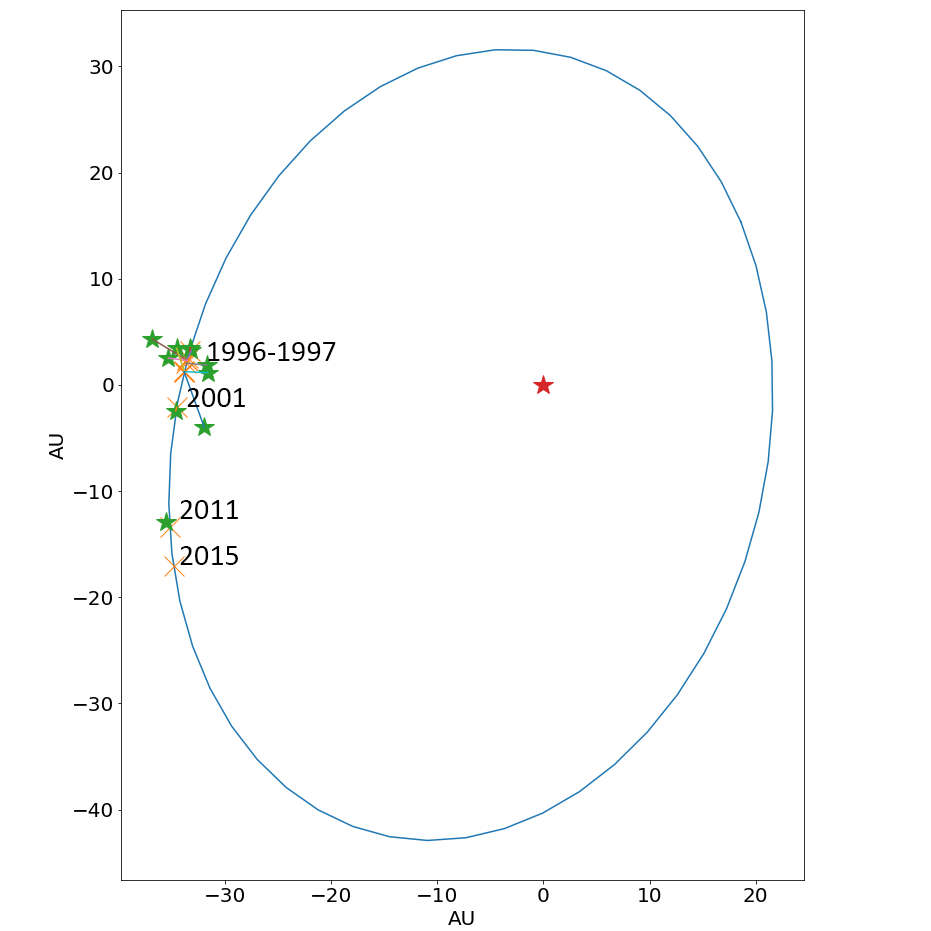}{0.4\textwidth}{d}
    \label{fig:orbit}
	}
	\caption{(a) Stokes $I$ image of FS Tau A, taken by Subaru/HiCIAO in H-band. Stars show the positions of the binary determined by PSF fitting.  (b) Averaged azimuthal profile of FS Tau A Stokes $I$ image within 100 pixels; the profile is drawn every 5$^\circ$. (c) \textit{Top}: radial profile of FS Tau A Stokes $I$ image along a position angle of 110/290$^\circ$ (blue dots). The lines show the fitted PSF profile of FS Tau Aa and Ab (orange), that of only FS Tau Aa (green), and that of only FS Tau Ab (red). \textit{Bottom}: residual map of the fitting; residuals were calculated as (observed value - fitted value)/fitted value. (d): The best-fit orbit of FS Tau A system. The red star is the position of FS Tau Aa, while the green stars are observed positions of FS Tau Ab since 1996, and the orange crosses are the predicted positions of FS Tau A.}
    \label{fig:profile}
\end{figure}

In the Stokes $I$ image shown in Figure~\ref{fig:fstaui}(a), a bright source and one faint structure extending to the southeast can be seen, indicating that this binary is not resolved well in our near-infrared observations due to the limited seeing correction. Therefore, to derive the separation $\rho$ and PA $\theta$ of the companion star Ab we tried the following methods. First, we determined the center of the Point Spread Function (PSF) via IRAF \textit{center} command, and then we plotted the azimuthal profile relative to this center to help determine the position angle (PA) of the binary. The averaged azimuthal profile, Figure~\ref{fig:azipro}(b) is drawn every 5$^\circ$ within 100 pixels. In this figure, the peak of the azimuthal profile appears at a PA of 110$^\circ$, indicating that the PA of the companion star is about 110$^\circ\pm2.5^\circ$.

To determine the separation of this binary, we plotted and fit the radial profile along the direction of the binary, i.e., a PA of 110$^\circ$, with an opening angle of 5$^\circ$. The PSF after AO correction should include one core structure and one halo structure, which can be approximately represented as the sum of two Gaussian functions. In this case, the PSF of the primary star FS Tau Aa $g_a$ can be expressed as:

\begin{equation}
g_a=a_ae^{-\frac{x^2}{2\sigma_h^2}}+sa_ae^{-\frac{x^2}{2\sigma_c^2}}
\end{equation}

Here, $a_a$, $\sigma_h$, $s$, and $\sigma_c$ represent the amplitude of the PSF halo, deviation of the halo, ratio of the amplitude of the core to that of the halo, and deviation of the core, respectively. For FS Tau Ab, we assume that its PSF has the same $s$, $\sigma_h$, and $\sigma_c$ as those of Aa, but a different amplitude $a_b$ and a different projected distance of $c$ pixels along PA=110$^\circ$ from Aa. As a result, its PSF can be written as:

\begin{equation}
g_b=a_be^{-\frac{(x-c)^2}{2\sigma_h^2}}+sa_be^{-\frac{(x-c)^2}{2\sigma_c^2}}
\end{equation}

Therefore, the total PSF profile can be expressed as:

\begin{equation}
G=g_a+g_b
\end{equation}

The fitting is performed by the SciPy \textit{curve\_fit} command, which employs the Levenberg-Marquardt method. The initial guesses of all parameters are 1. The PSF fitting results as well as their standard deviation errors are listed in Table 1 ($r^2$ is the coefficient of determination) and shown in Figure~\ref{fig:radpro}(c), they do not have significant changes when we change the initial guesses of the parameters. To better describe how our results fit the observations, we also show the residual map. From the residual map, it can be seen that the relative residuals, calculated as (observed value - fitted value)/fitted value, of the fitting are smaller than 0.1 within a radius of $\sim$90 pixels, and are extremely small in the direction of Ab, indicating that our fitted PSF profile can be regarded as a good approximation of the observed profile. From the curve fitting result, we derived separation $c=-28.53$ pixels, corresponding to 0.$\arcsec$271. As for the uncertainties of the binary separation, we consider the pixel scale, the accuracy of the image registration, and the errors of the fitting. The standard deviation error from the fitting is about 0.13 pixel. Also the radius is calibrated from the brightness peak, for simplicity we assume that the star Aa is at the same place (zero point) as the brightness peak, however there is a $\sim$0.9 pixel offset between them in the fitting result. It is hard to make the fitted brightness peak just at the zero point so we regard it as an error. In summary, we estimate that the astrometric error of the separation should not be larger than about 1 pixel or 9.5 mas. 

According to the orbit derived by \citet{Tamazian2002}, the predicted position of FS Tau Ab should be at $\rho=0.\arcsec224\pm 0.\arcsec02$ and $\theta=112.4^\circ \pm 18^\circ$, which is not consistent with our result. Therefore, we tried to fit a new orbit. The MCMC algorithm of the code \textit{orbitize!} \citep{Blunt2019} is used to fit the orbit. Our data, as well as the data taken from the year 1996 to 2001 in \citet{Tamazian2002} are used for fitting. The 1989 year data used by \citet{Tamazian2002} were derived from lunar occultation, considering that this method have larger uncertainties, we do not use these results in our fitting. We set the parallax $\pi$ of the binary to be 7.143$\pm$0.001 mas. As for the system mass $M_{sys}$, we found out that the best solutions with the lowest chi-squared values (the chi-squared values of the orbits are calculated via $\sum (O-E)^2/E$, here $O$ and $E$ represent the observed positions and expected positions of the observed epoch, respectively) during the test runs tend to have a mass around 1.5$M_\odot$ so we set the initial mass at 1.5$\pm$0.25 $M_\odot$. The Gaussian prior $log(p(x|(\delta,\mu))\propto\frac{x-\mu}{\delta}$ is used for the parallax and system mass. As for other parameters, we do not set initial values. The semi-major axis $a$ uses Log Uniform prior $p(x)\propto1/x$, the inclination $i$ uses Sine prior $p(x)\propto sin(x)$, and the other parameters use Uniform prior.

\begin{figure}[htbp!]
\figurenum{2}
\centering    
\includegraphics[scale=0.35]{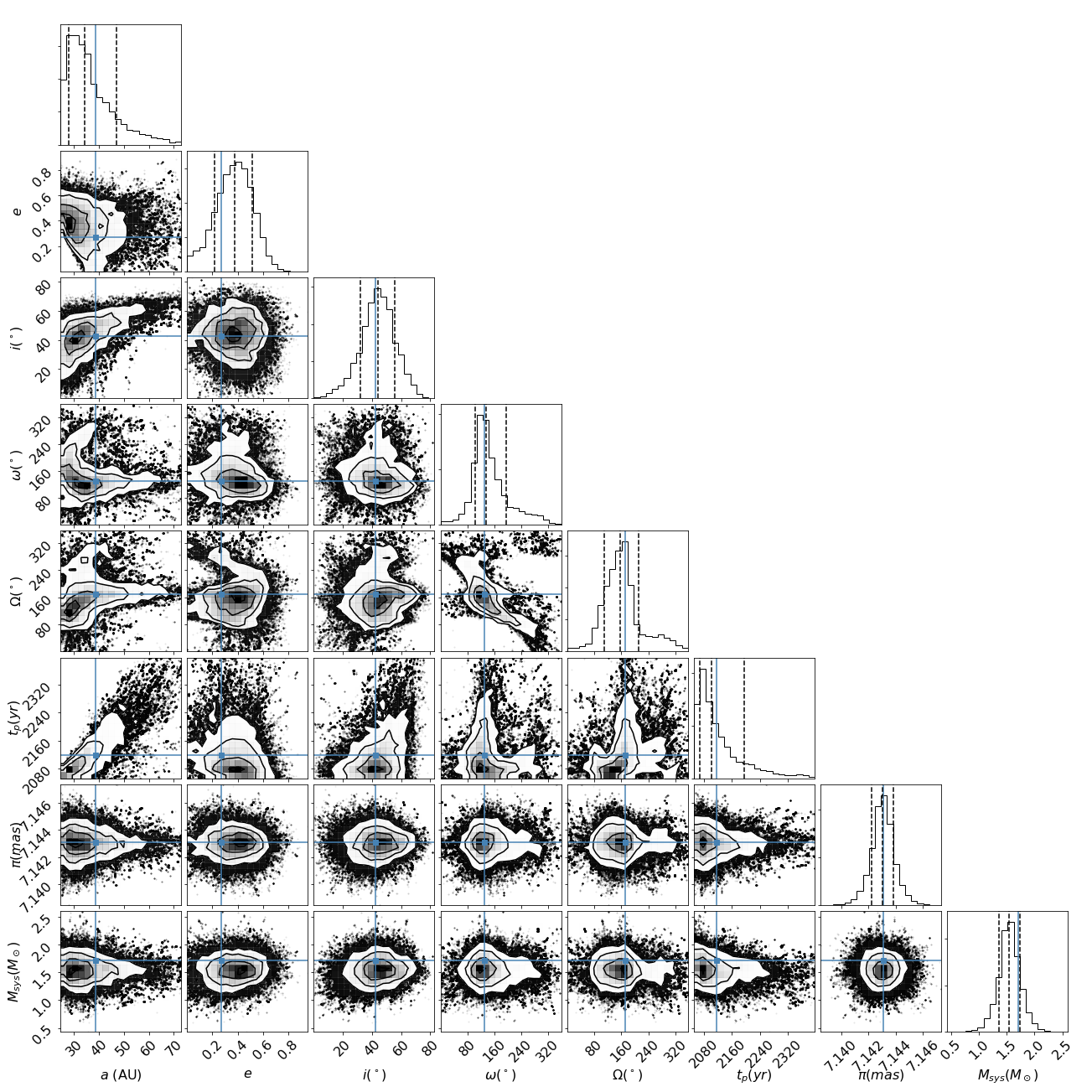}
\caption{Corner map of the MCMC fitting result. The dashed lines show the 0.16, 0.5 and 0.84 percentile of the MCMC results, while the blue lines indicate the position of the best-fit ($\chi^2_{min}$) value.}
\label{fig:corner}
\end{figure}

The fitting result is shown in Table 2, which is a sample including 1000000 orbits following the convergence. The median, lower and upper values correspond to 0.5, 0.16 and 0.84 percentile of the results, respectively, while the $\chi^2_{min}$ indicates the best-fit result among the orbits, which has a $\chi^2_{min}$ value of about 1.88, and is drawn in Figure~\ref{fig:orbit} (d). From this image, it can be seen that this result fits the observation results of 2001 and 2011 quite well. The corner map is also shown in Figure~\ref{fig:corner}.

\begin{deluxetable}{ccc}\tablecaption{PSF fitting results and derived binary positions}
\tablecolumns{3}
\tablenum{1}
\tablewidth{0pt}
\tablehead{
\colhead{Parameter} &
\colhead{Value} &
\colhead{Error} 
}
\startdata
$a_a$ (ADU) & 4650 & 43  \\
$\sigma_h$ (pixel) & 41.52 & 0.16 \\
$s$ & 2.79 & 0.03  \\
$\sigma_c$ (pixel) & 13.10 & 0.06 \\
$a_b$ (ADU) & 1210 & 13  \\
$c$ (pixel) & -28.53 & 0.13(1)  \\
$r^2$ & 0.99833 & - \\ \hline
$\rho(\arcsec)$ & 0.271 & 0.0095 \\
$\theta(^\circ)$ & 110 & 2.5 \\
\enddata
The errors are from the standard deviation errors. While the error in the colon for $c$ is the error after combining other possible uncertainties.
\end{deluxetable}

\begin{deluxetable}{ccccc}\tablecaption{Orbit fitting results}
\tablecolumns{5}
\tablenum{2}
\tablewidth{0pt}
\tablehead{
\colhead{Parameter} &
\colhead{Median} &
\colhead{Lower} &
\colhead{Upper} &
\colhead{$\chi^2_{min}$}
}
\startdata
$a$ (AU) & 34.3 & 28.0 & 46.9 & 38.5 \\
$e$ & 0.374 & 0.220 & 0.512 & 0.269 \\
$i(^\circ)$ & 44.3 & 32.2 & 55.5 & 42.6  \\
$\omega(^\circ)$ & 135 & 103 & 197 & 130 \\
$\Omega(^\circ)$ & 158 & 110 & 211 & 170  \\
$t_p(yr)$ & 2101.39 & 2069.03 & 2195.44 & 2117.69  \\
$\pi(mas)$ & 7.1430 & 7.1422 & 7.1438 & 7.1431 \\
$M_{sys}(M_{sol})$ & 1.54 & 1.36 & 1.74 & 1.71 \\
\enddata
\end{deluxetable}

\subsubsection{Circumstellar Structures}

\begin{figure}[ht!]
	\figurenum{3}
    \gridline{
    \fig{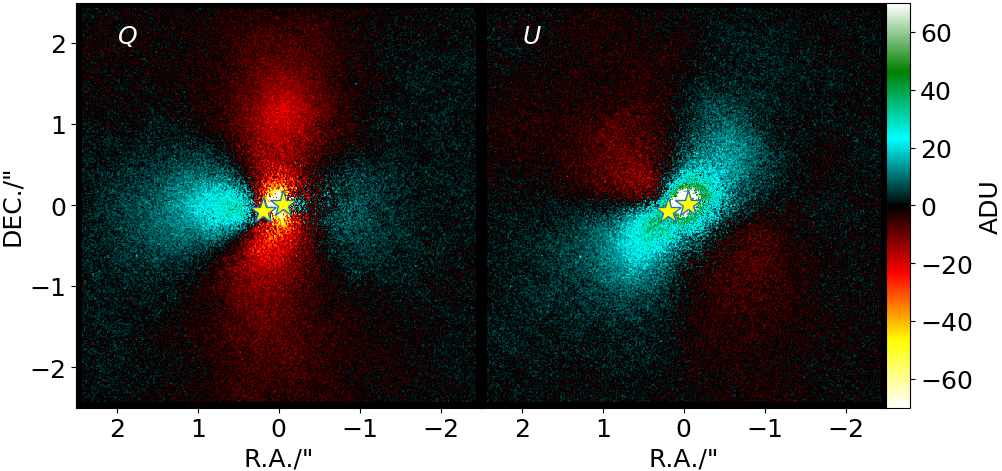}{0.8\textwidth}{a}
    \label{fig:stokes}
	}
	\gridline{
    \fig{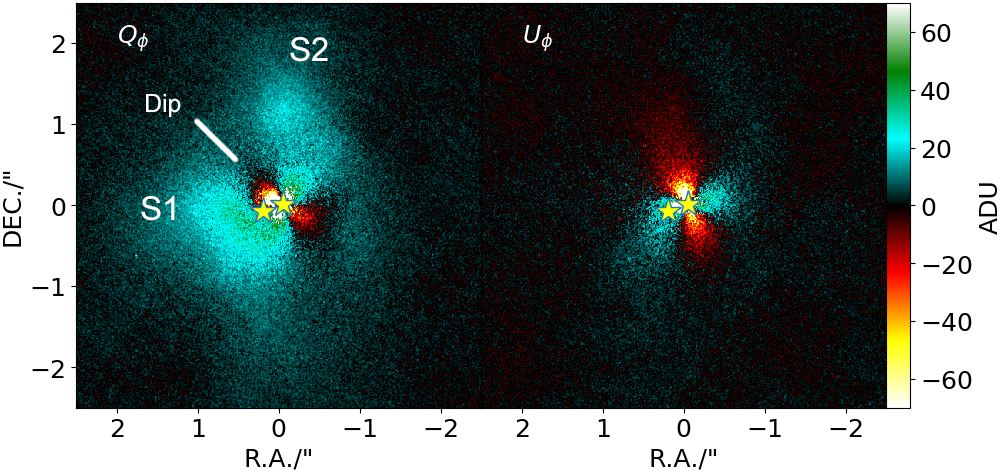}{0.8\textwidth}{b}
    \label{fig:rstokes}
	}
	\gridline{
    \fig{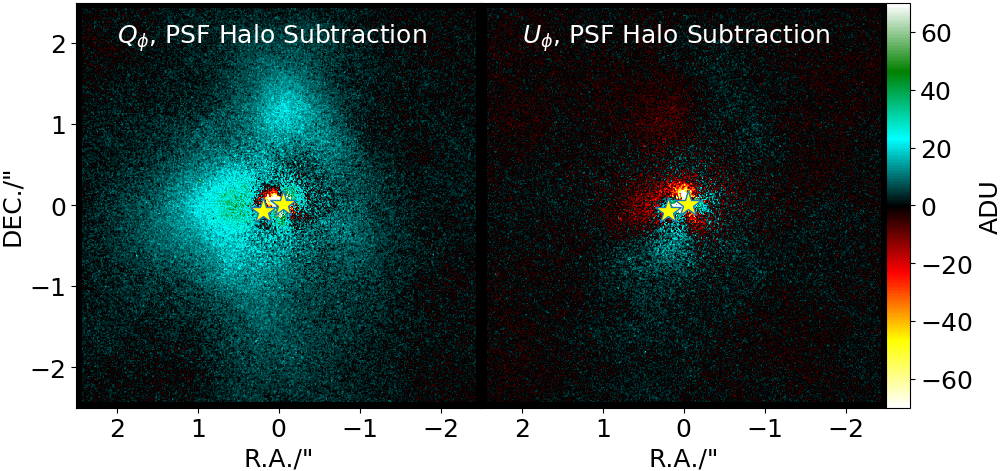}{0.8\textwidth}{c}
    \label{fig:rstokeshs}
	}
	\caption{(a) Stokes $Q$ and $U$ images of FS Tau A. The stars show the position of stars as Figure~\ref{fig:fstaui}(a). (b) derived from the Stokes $Q_\phi$ (b) and $U_\phi$ (c) images of FS Tau A. (c) the same but after PSF halo Subtraction.}
    \label{fig:nirresult}
\end{figure}

We calculated the radial Stokes images relative to the brightness peak of the Stokes $I$ image using the methods suggested by \citet{Avenhaus2014}, based on the Stokes $Q$ and $U$ (Figure~\ref{fig:stokes}(a)). In the $Q_\phi$ radial Stokes image shown in Figure~\ref{fig:rstokes}(b), it can be seen that this binary system is accompanied by nebular-like structures. Two symmetric, butterfly-like negative regions along a PA of about 55$^\circ$ and a radius of $\sim0.\arcsec5$ can be seen, indicating that the polarization directions in these regions are along the radial direction. Since the instrumental polarization correction has been done via the method described by \citet{Joos2008}, other mechanisms should be responsible for this butterfly-pattern. One possibility is that it is caused by the uncorrected PSF halo of the stars \citep[e.g.,][]{Oh2016}.

To test this we tried to generate the PSF halo for Stokes $Q$ and $U$ images using the method similar to \citet{Oh2016}: firstly we derived the brightness ratio between Stokes $Q/U$ images and Stokes $I$ image, by calibrating their brightness between radius 13.5-105 pixels relative to the brightness center position of Stokes $I$ image, to estimate the brightness of the PSF halos in Stokes $Q$ and $U$ images (about -0.3\% and 0.5\% relative to the Stokes $I$ image, respectively). Then we generated the PSF halos based on the profile and brightness ratio we derived above, and we subtracted them from the origin Stokes $Q$ and $U$ images to get the PSF halo subtracted Stokes $Q$ and $U$ images. Finally we re-calculated the radial Stokes images after PSF halo subtraction (Figure~\ref{fig:rstokeshs}(c)). From the $Q_\phi$ image in Figure~\ref{fig:rstokeshs}(c), we notice that the negative regions become smaller, however much artifacts are also introduced into the image, especially the ``dip", which was previously observed by \citet{Hioki2011}, disappears in the PSF halo-subtraction image. Toward this, we suggest that uncorrected PSF halo may not be the mechanism causing this butterfly pattern. Considering the extinction of FS Tau A is high ($A_V\sim$5, \citet{Kraus2009}),  such as the the multiple scattering due to the dust in front of the stars, could contribute to that. Following observations can tell us more about it and this time we will mainly focus on the other detected structures, i.e., structures outside $\sim0.\arcsec5$ from the star.  

\begin{figure}[ht!]
	\figurenum{4}
    \gridline{
    \fig{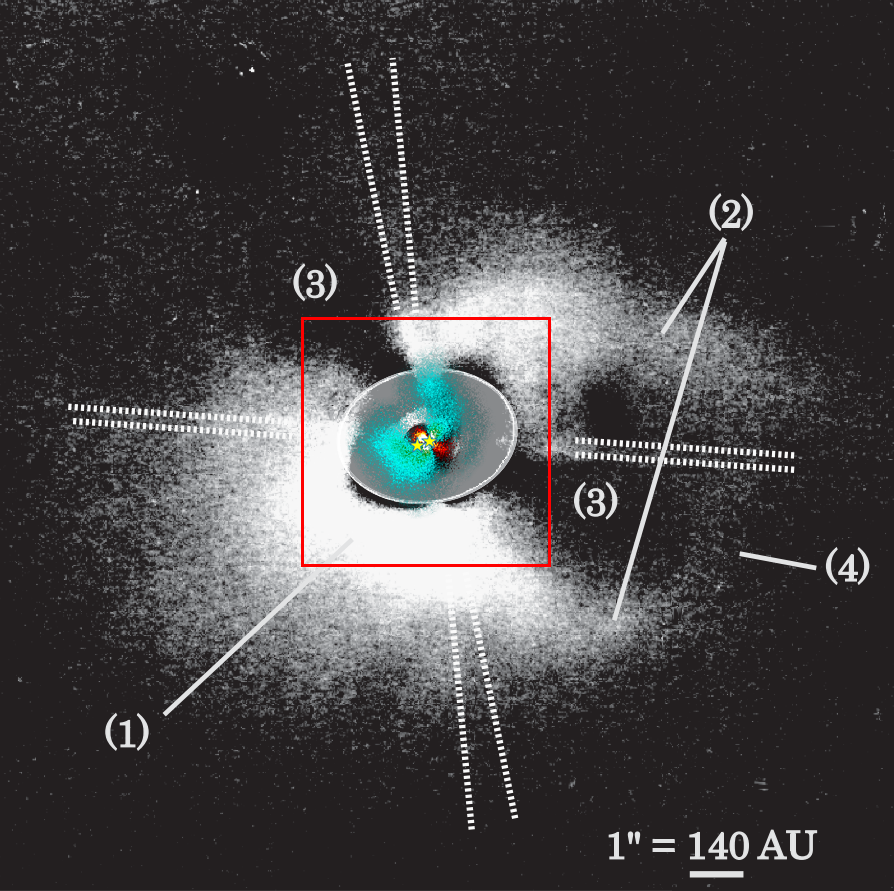}{0.4\textwidth}{a}
    \label{fig:composite}
    \fig{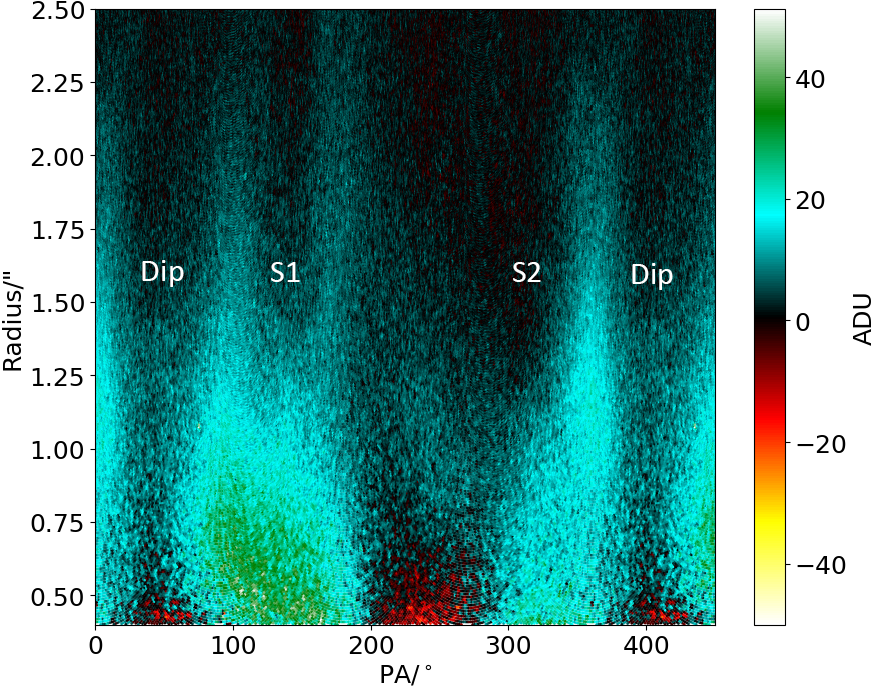}{0.5\textwidth}{b}
    \label{fig:radialmap}
	}
	\caption{(a) The FS Tau A image of combining our result and \citet{Hioki2011}. The $5\arcsec\times5\arcsec$ red square indicates the area we see from HiCIAO. (b) Azimuthal mapping of Stokes $Q_\phi$ image with radius ranging from 0.$\arcsec$4 to 2.$\arcsec$5. The x-axis is drawn from 0$^\circ$ to 450$^\circ$ to clearly show the S2 structure.}
    \label{fig:fstau}
\end{figure}

In the composite image with the result of \citet{Hioki2011}, Figure~\ref{fig:composite}(a), it can be seen that we successfully revealed the structures within the $1.\arcsec3\times1.\arcsec7$ area where \citet{Hioki2011} did not resolved. In the $Q_\phi$ image, one structure, labeled ``S1", was detected at the 5-6$\sigma$ level in the southeast region (signal is from the $Q_\phi$ image, noise level was estimated from the standard deviation of the $U_\phi$ image at the same place). The structure starts from the south of the binary, and then turns to the northeast. From the azimuthal mapping image shown in Figure~\ref{fig:radialmap} (b), its main part ranges from PA $\sim$70-200$^\circ$ and extends to at least $\sim1.\arcsec5$ from the stars. Another structure, labeled ``S2", was detected at the $4-5\sigma$ level in the north. From Figure \ref{fig:composite}(a) it seems to be connected with the (2) structure of \citet{Hioki2011}. It ranges from PA $\sim$300-380$^\circ$ and extends to at least $\sim1.\arcsec7$ from the stars. There is a dip centered on PA $\sim50^\circ$ between the S1 and S2 structures, whose position is consistent with the northeast dip (3) detected by \citet{Hioki2011} (Figure~\ref{fig:composite}(a)).

\subsection{Sub-millimeter Observations}

\begin{figure}[htbp!]
\figurenum{5}
\centering    
\includegraphics[scale=0.7]{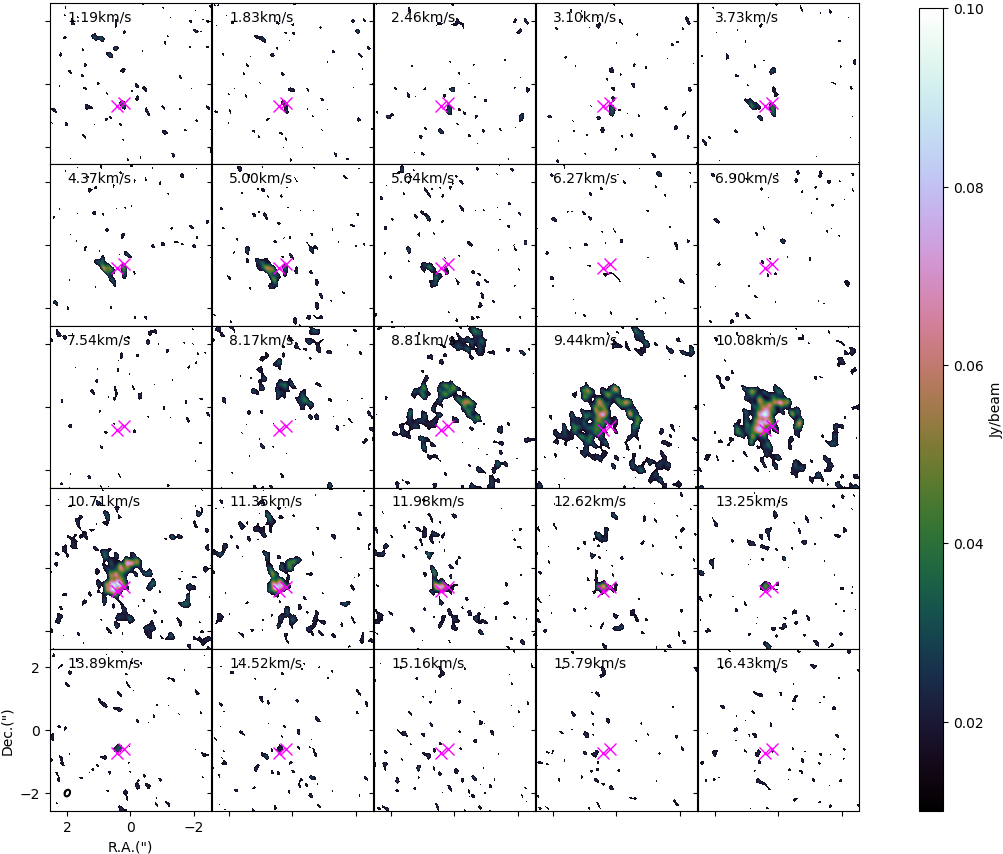}
\caption{ALMA CO 2-1 channel map of FS Tau A. LSRK velocities are shown in the upper-left corners of channels. The beam size is shown as an open circle in the lower-left corner of the bottom left (19.89 km/s) channel.}
\label{fig:fstauco21ch}
\end{figure}

From the channel map of FS Tau A CO ($J$=2-1) image taken by ALMA shown in Figure~\ref{fig:fstauco21ch} (beam size: 0.$\arcsec16\times0.\arcsec$22; 
PA: 26.24$^\circ$), it can be seen that emissions were detected in channels 1-8 (local standard of rest kinematic (LSRK) velocity: 1.19-5.64 km/s) and 12-25 (LSRK velocity: 8.17-16.43 km/s). 
No emission was detected in channels with LSRK velocities of 6.27, 6.90, and 7.54 km/s, corresponding to solar system barycentric radial velocities of 15.68, 16.31, and 16.94 km/s, respectively, indicating that two independent components were detected in the ALMA CO ($J$=2-1) image. 
No calibration of FS Tau A's radial velocity was made. However, we noticed that its two nearby T Tauri stars, namely BD +26 718B and IP Tau, with separations of about 44 and 46 arcmin from FS Tau A, respectively, have barycentric radial velocities of 16.23 and 16.24 km/s, respectively \citep{Nguyen2012}. In this case, we suggest that the radial velocity of FS Tau A is not likely to be far from these values. Therefore, it is reasonable to assume that the heliocentric radial velocity of FS Tau A is 16.3 km/s, corresponding to an LSRK velocity 6.90 km/s. The detected structures can then be classified as blue- and red-shifted parts.

\begin{figure}[ht!]
	\figurenum{6}
    \gridline{
    \fig{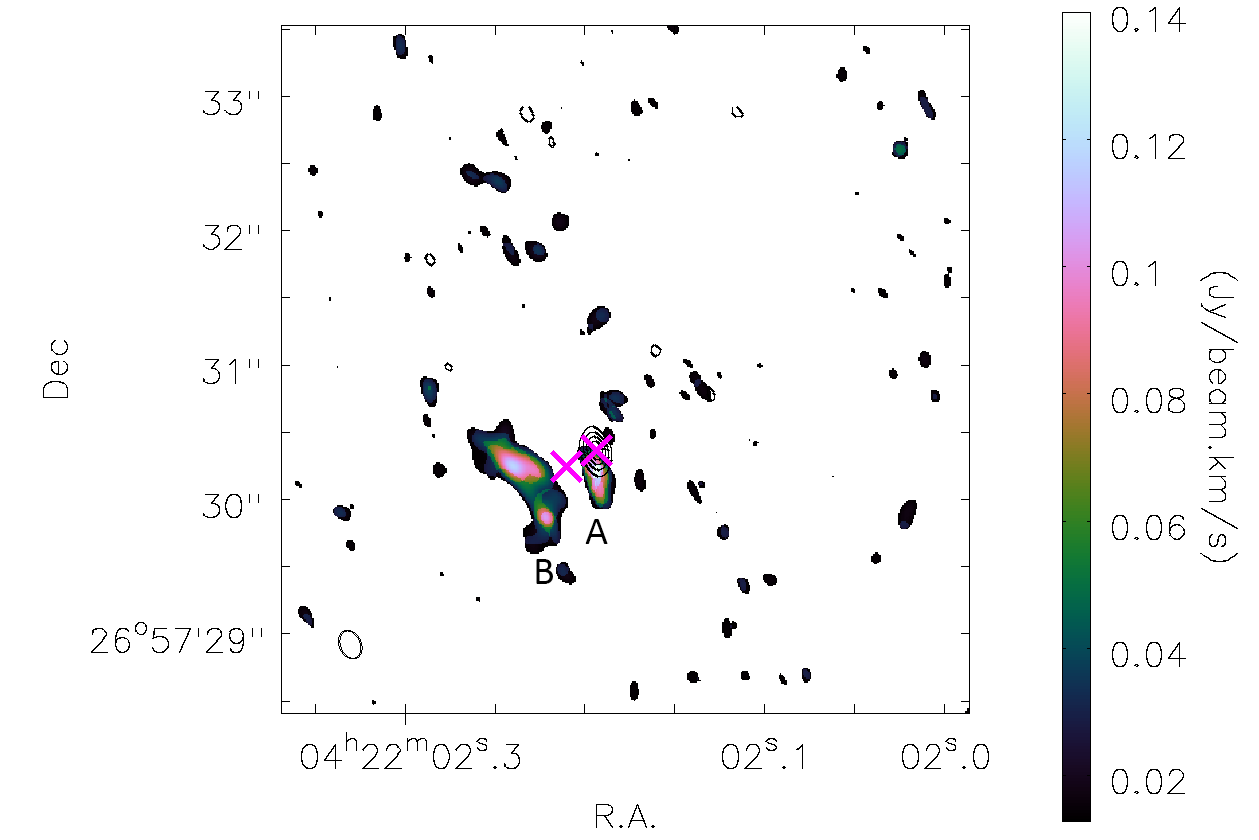}{0.5\textwidth}{a}
    \label{fig:fstau0comp1}
    \fig{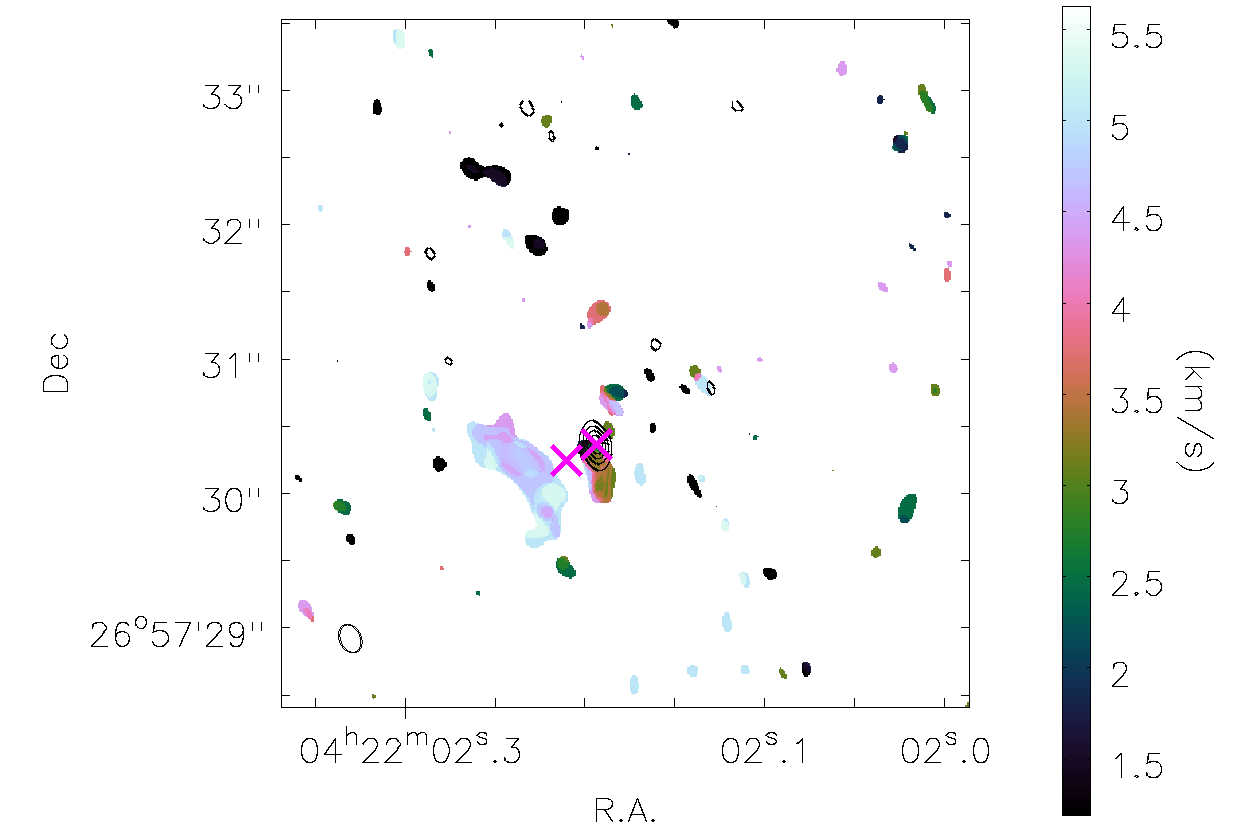}{0.5\textwidth}{b}
    \label{fig:fstau1comp1}
    }
    \gridline{
    \fig{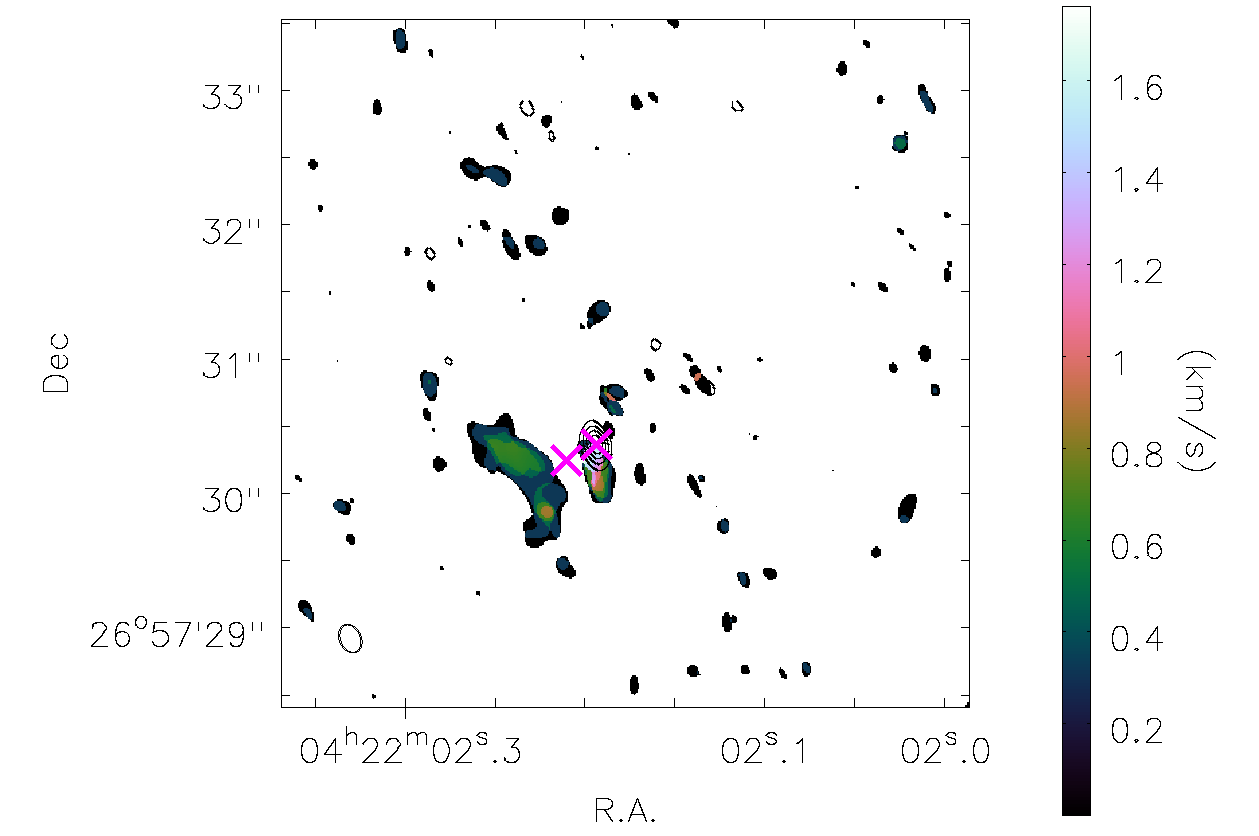}{0.5\textwidth}{c}
    \label{fig:fstau2comp1}
    \fig{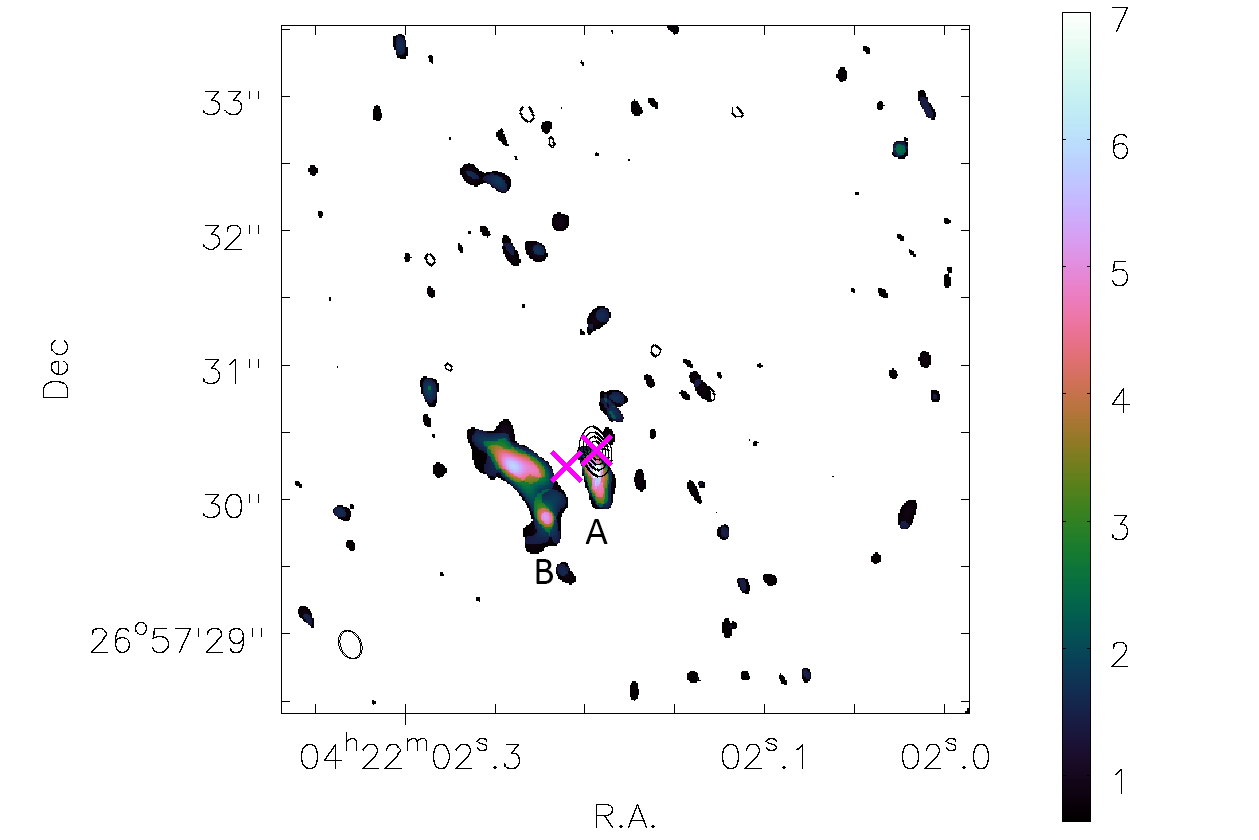}{0.5\textwidth}{d}
    \label{fig:fstausnmapcomp1}
	}
	\caption{ALMA CO 2-1 blue-shifted component images, corresponding to velocities of 1.19-5.64 km/s. (a) ALMA CO 2-1 moment 0 image of FS Tau A blue-shifted component. (b) ALMA moment 1 image of FS Tau A blue-shifted component. (c) ALMA moment 2 image of FS Tau A blue-shifted component. (d) S/N map of (a). The black contours in the images represent the continuum emission image showing the position of the central stars, showing 0.2, 0.4, 0.6 and 0.8 of the peak intensity 2.08 mJy/beam. The red crosses show the positions of FS Tau Aa and Ab derived from our near-infrared observations, respectively, assuming that the peak of the continuum emission corresponds to Aa.}
    \label{fig:fstaualmacomp1}
\end{figure}

We integrated these two independent components individually. The blue-shifted component was integrated from channels 1-8 (1.19-5.64 km/s); its moment 0 (integrated flux), 1 (velocity fields), 2 (velocity dispersions) images and the signal noise ratio map are shown in Figure~\ref{fig:fstaualmacomp1}. In the moment 0 image, two obvious structures can be seen. One structure, labeled structure A, extends from the star Aa to about 0.$\arcsec$4 or 56 AU at the 3$\sigma$ level (1$\sigma \sim$0.02 Jy/beam$\cdot$km/s) with a PA of about 190$^\circ$; its brightest part is about 7$\sigma$. In the moment 1 map, it can be seen that the south part of structure A generally has a velocity of about 3.5 km/s, and at the position near the star Aa, there seems to be a velocity gradient. In the moment 2 map, there is clearly a larger velocity dispersion of about 1.6 km/s near the star Aa. Another structure, a bar-like structure labeled structure B, is located at about 0.$\arcsec$15 or 21 AU from the star Ab. It has two cores: a large one in the northeast and a small one in the southwest. Both cores were detected at about 5$\sigma$ with a separation of about 0.$\arcsec$7 or 98 AU and a PA of about 30$^\circ$. The connection between these two cores was barely detected ($\sim2\sigma$). In the moment 1 map, structure B has a velocity of about 5 km/s, and the two cores have slightly larger velocity dispersions ($\sim$0.6 km/s for the north core and $\sim$0.8 km/s for the south core) than that of their connecting part ($\sim$0.3 km/s) in the moment 2 map.

\begin{figure}[htbp!]
\figurenum{7}
\gridline{
    \fig{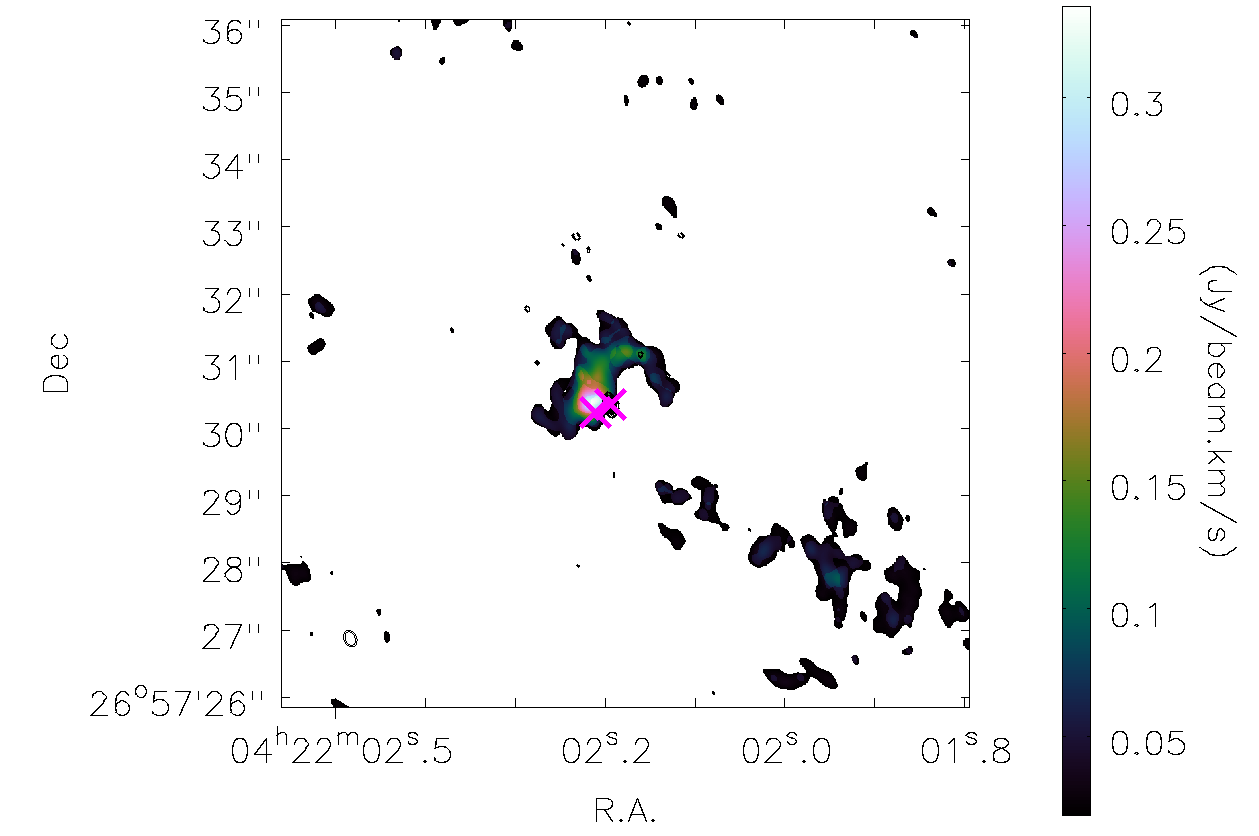}{0.5\textwidth}{a}
    \label{fig:fstaualma0}
    \fig{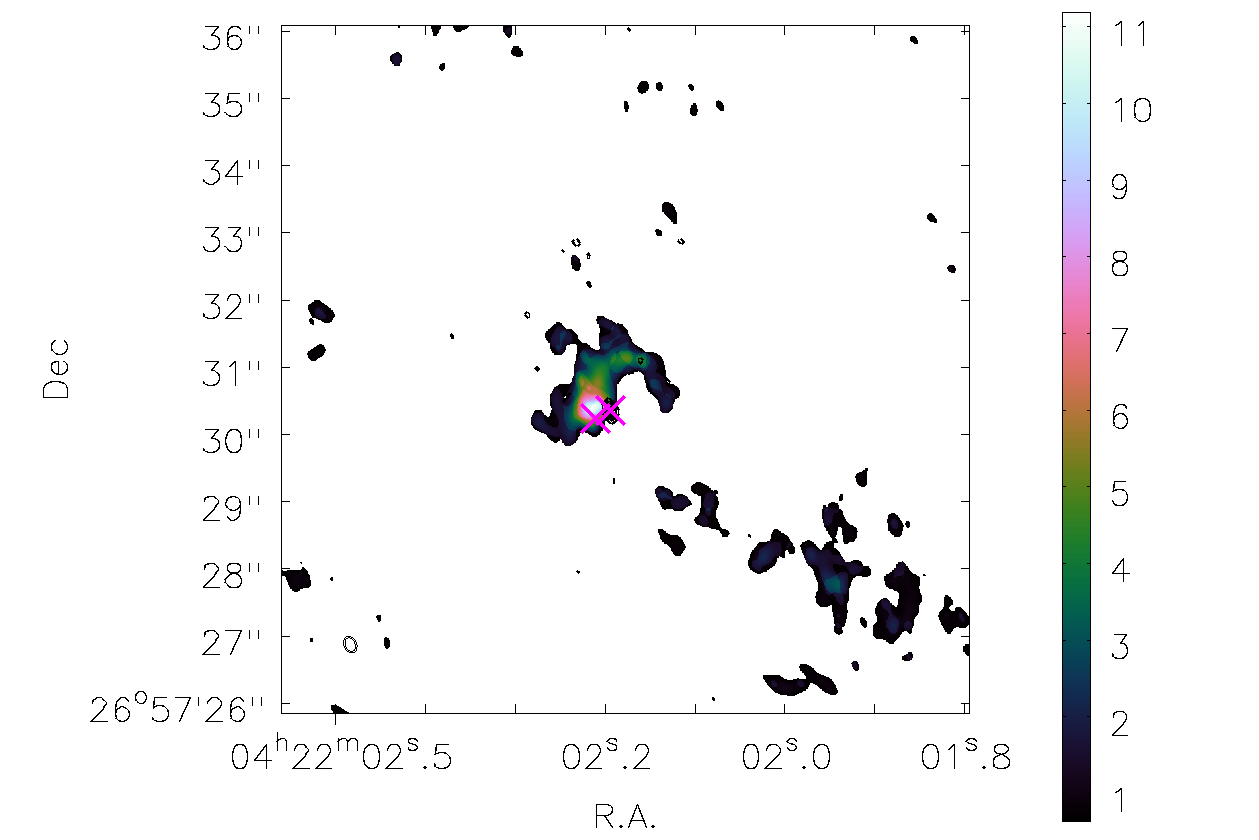}{0.5\textwidth}{b}
    \label{fig:fstaualma0_sn}
	}
	\caption{(a) ALMA CO 2-1 red-shifted component moment 0 images, corresponding to velocities of 8.17-16.43 km/s, with a large (10$\arcsec\times$10$\arcsec$) field of view. (b) S/N map of (a).}
\label{fig:fstau_large}
\end{figure}

The red-shifted component was integrated from channels 12-25 (8.17-16.43 km/s). Before analyzing the central structures, it is necessary to mention that some structures extending to about 5.$\arcsec$5 (770 AU) were barely detected in the southwest (Figure~\ref{fig:fstaualma0}(a)). These structures seem to be located at the southwest boundary of the "2" and "3" structures in the image reported by \citet{Hioki2011}, and have an LSRK velocity of about 9.5 km/s and a small velocity dispersion (0.3 km/s). However, they are too faint ($\sim2-3\sigma$, $1\sigma\sim$0.03 Jy/beam$\cdot$km/s, Figure~\ref{fig:fstaualma0_sn}(b)) in this ALMA observation; further observation is needed for a detailed discussion. Here, we focus on the structures detected within 5$\arcsec$ from the stars.

\begin{figure}[ht!]
	\figurenum{8}
    \gridline{
    \fig{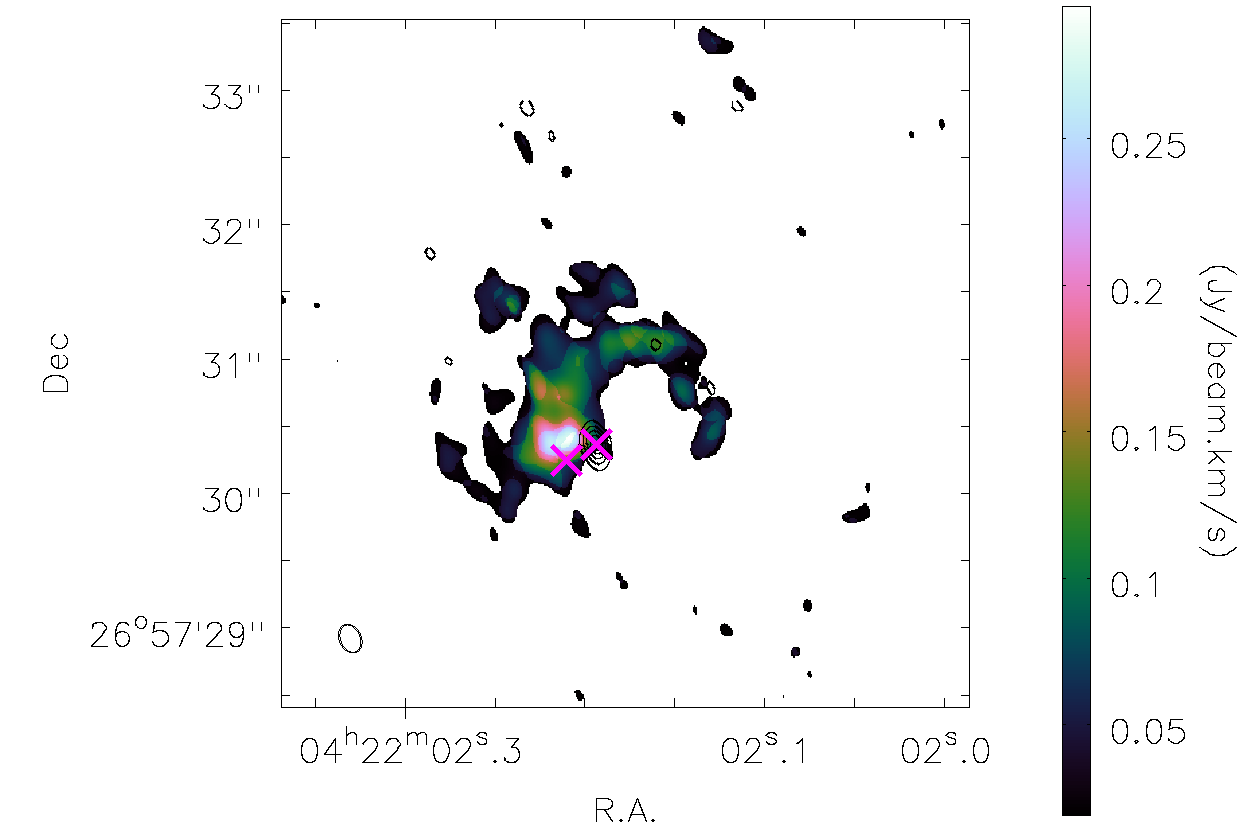}{0.5\textwidth}{a}
    \label{fig:fstau0comp2}
    \fig{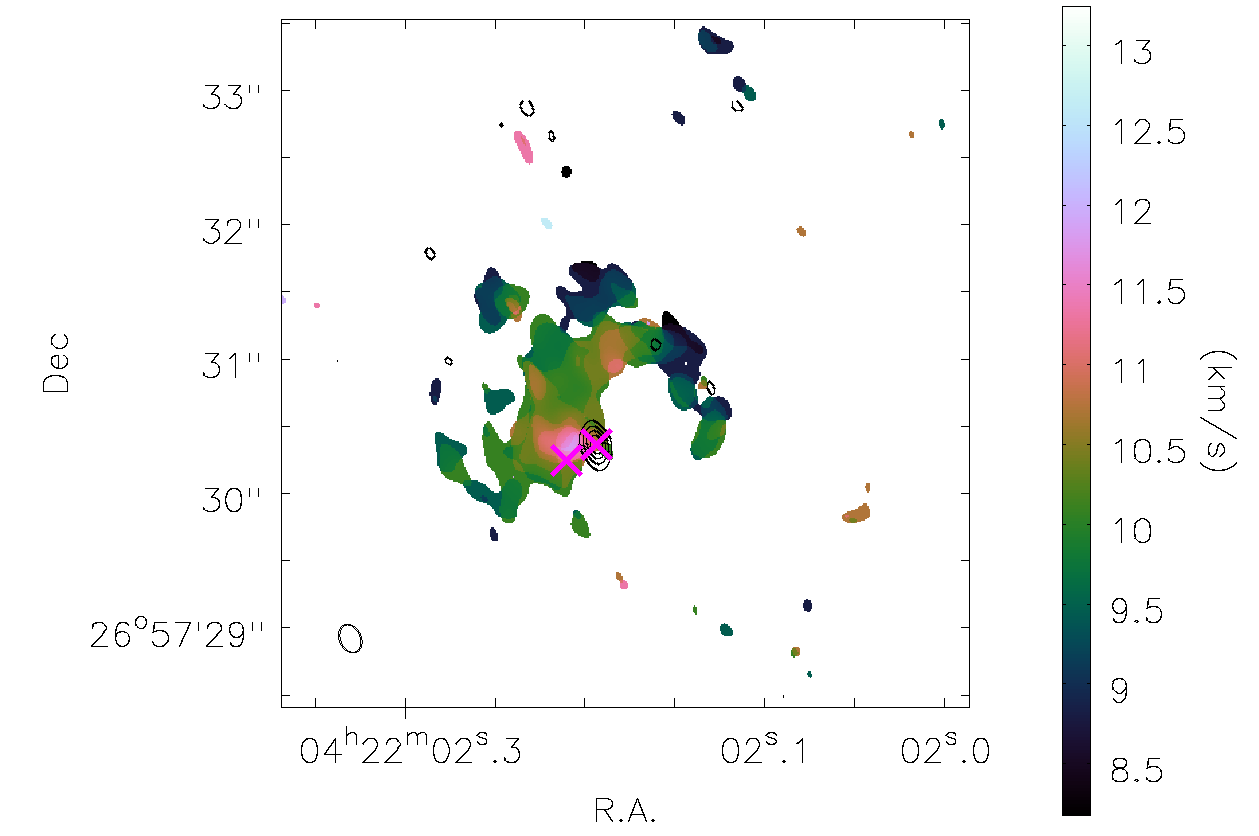}{0.5\textwidth}{b}
    \label{fig:fstau1comp2}
    }
    \gridline{
    \fig{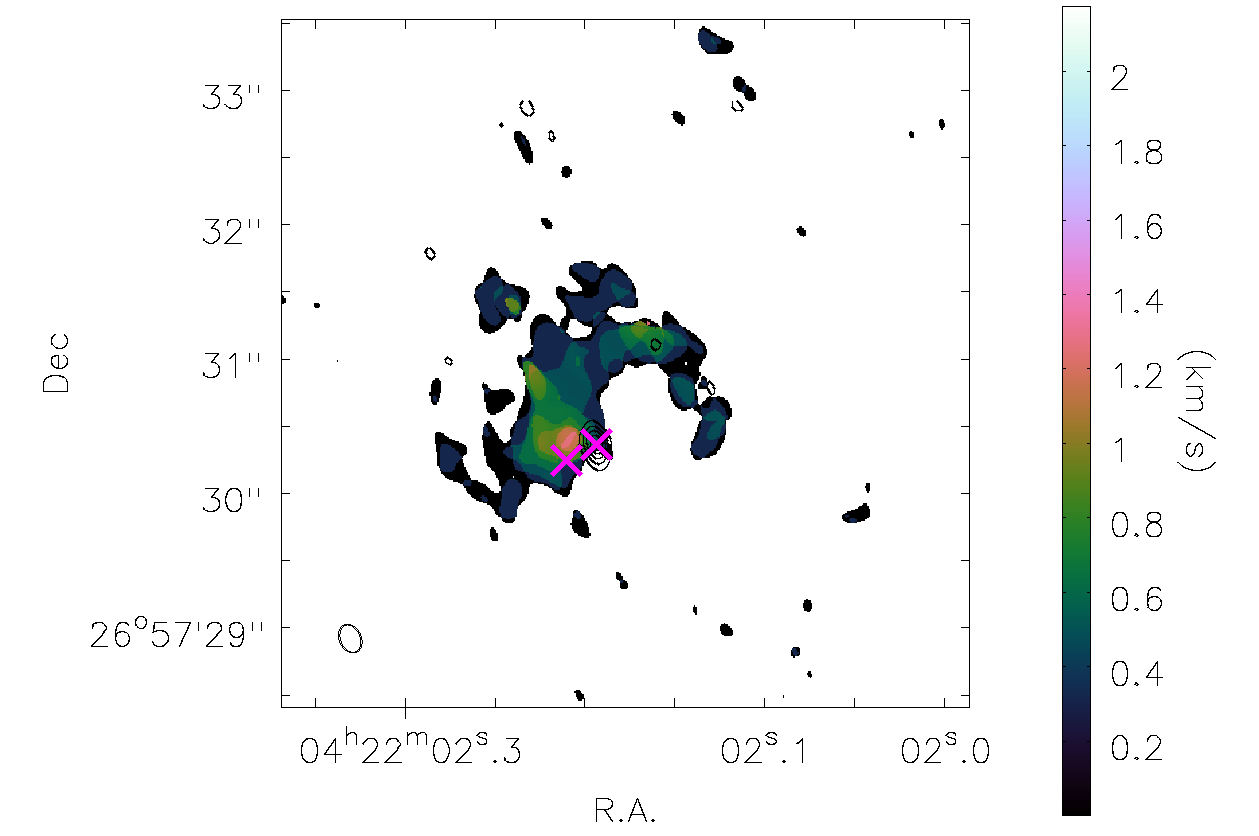}{0.5\textwidth}{c}
    \label{fig:fstau2comp2}
	}
	\caption{ALMA CO 2-1 images of red-shifted component central region, corresponding to velocities of 8.17-16.43 km/s. (a) ALMA CO 2-1 moment 0 image of FS Tau A red-shifted component. (b) ALMA moment 1 image of FS Tau A red-shifted component. (c) ALMA moment 2 image of FS Tau A red shifted component. Black contours represent ALMA 1.3-mm continuum image, showing 0.2, 0.4, 0.6 and 0.8 of the peak intensity 2.08 mJy/beam.}
    \label{fig:fstaualmacomp2}
\end{figure}

The moment 0, 1, and 2 images of the red-shifted component within 5$\arcsec$ are shown in Figure~\ref{fig:fstaualmacomp2}. In the center of the moment 0 image, there is a very bright structure ($\sim9\sigma$) close to the two stars in the northeast, and in the moment 1 and 2 maps, it can be seen that this structure has larger velocity ($\sim$12 km/s) and velocity dispersion ($\sim$1.3 km/s) than those of the other surrounding structures.
There is also an arm extending first to the north to $\sim0.\arcsec7$ (98 AU), and turning to the west to $\sim0.\arcsec6$ (84 AU) from the turning point at generally $>4\sigma$. There is also some emission at $\sim2\sigma$ extending from the end of the arm to the south, which may indicate that this arm finally turns to the south. 

\begin{figure}[ht!]
	\figurenum{9}
    \gridline{
    \fig{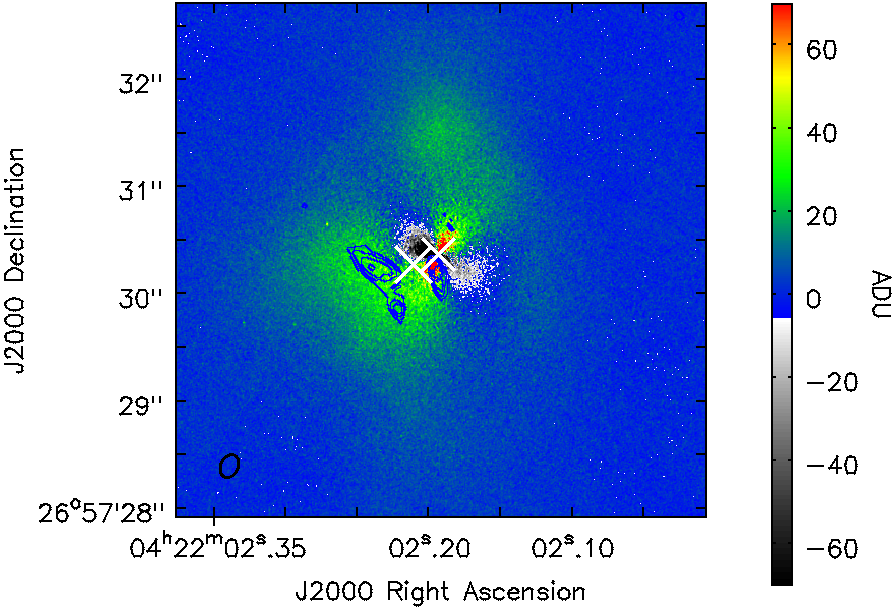}{0.5\textwidth}{a}
    \label{fig:comp1overpi}
    \fig{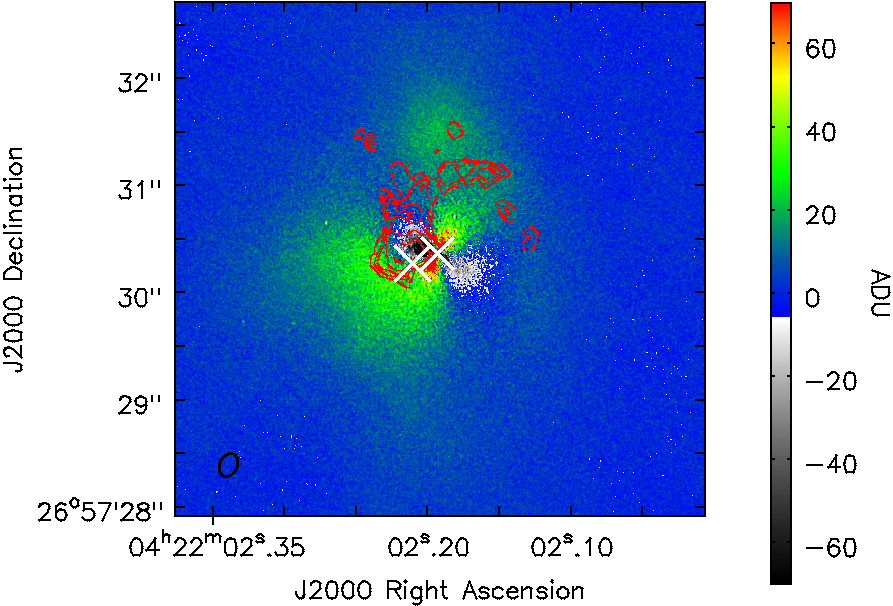}{0.5\textwidth}{b}
    \label{fig:comp2overpi}
	}
	\caption{(a): $Q_\phi$ image (color scale) with ALMA CO 2-1 blue-shifted component image (contour) overplotted. Contours are 0.03822, 0.06374, 0.08926, 0.1148 Jy/beam; (b): $Q_\phi$ image (color scale) with ALMA CO 2-1 red-shifted component image (contour) overplotted. Contours are 0.0606, 0.0883, 0.116, 0.144, 0.227 Jy/beam.}
    \label{fig:fstauoverplot}
\end{figure}

The near-infrared observations and the sub-millimeter CO ($J$=2-1) observations are compared in Figure~\ref{fig:comp1overpi}. Figure~\ref{fig:comp1overpi}(a) shows a comparison of the blue-shifted component and the near-infrared observations ($Q_\phi$ image). It can be seen that the blue-shifted component structures generally correspond to the S1 structure observed in the near-infrared band. We can see that structure B in the sub-millimeter band is consistent with the southeast part of structure S1 in the near-infrared band. As for Structure A, it is smaller than the butterfly pattern ($\sim0.\arcsec5$), but from its shape it seems to correspond to the S1 part extending from the star FS Tau Aa. 

A comparison between the red-shifted structures and the near-infrared observations is shown in Figure~\ref{fig:comp2overpi}(b). The red-shifted component arm first extends to the north, and then turns to the west. From Figure~\ref{fig:comp2overpi}(b), it can be seen that the position of the north-extending part lies slightly east of the S2 structure detected in the near-infrared observations. However, the west-extending part of the red-shifted component arm overlaps the S2 structure, indicating that there is some relationship between the structures detected in the different bands. Future observations can tell us more about this. 

\section{Discussion}

\subsection{Misaligned Inner Circumbinary Disk?}

We have shown that the near-infrared polarimetry observations and CO ($J$=2-1) observations resolved some structures around FS Tau A. In this section, we briefly discuss these structures. 

\begin{figure}[htbp!]
\figurenum{10}
\gridline{
    \fig{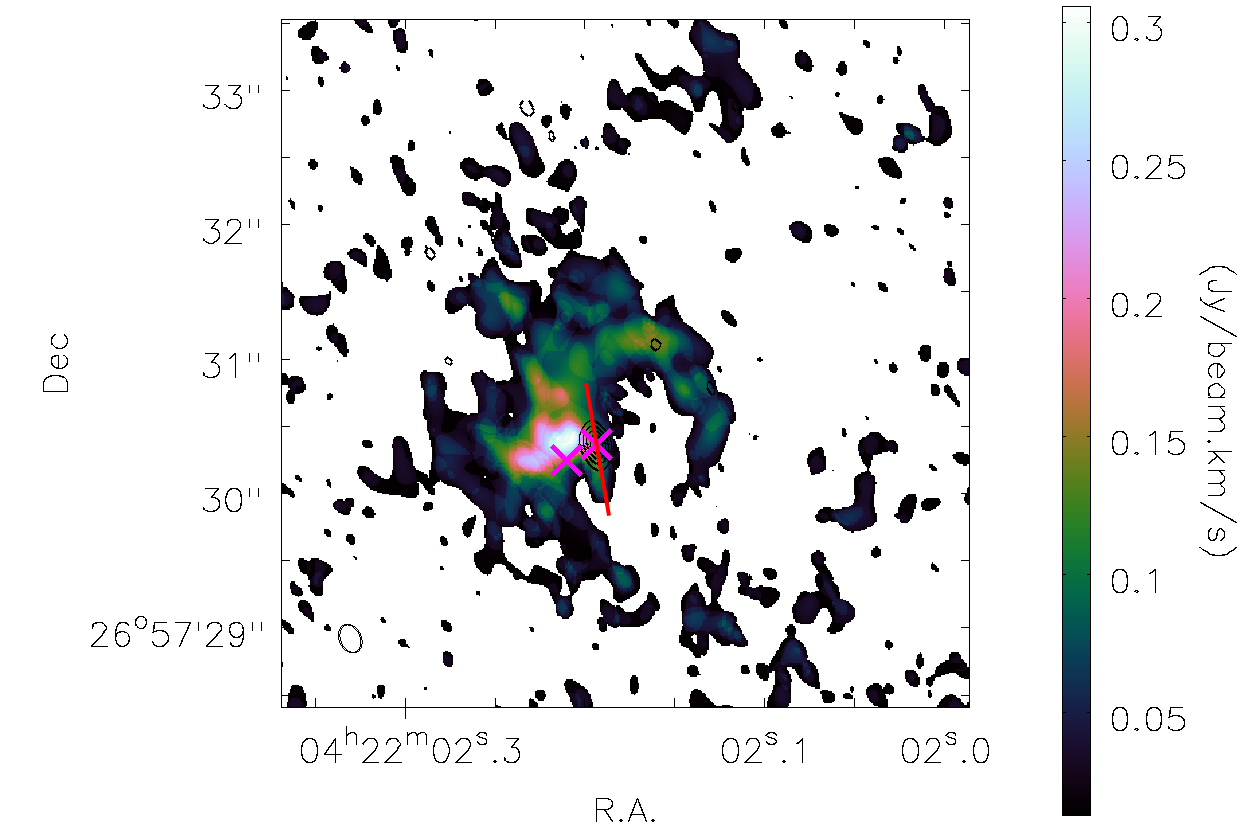}{0.6\textwidth}{a}
    \label{fig:fstaualma0_all}
    \fig{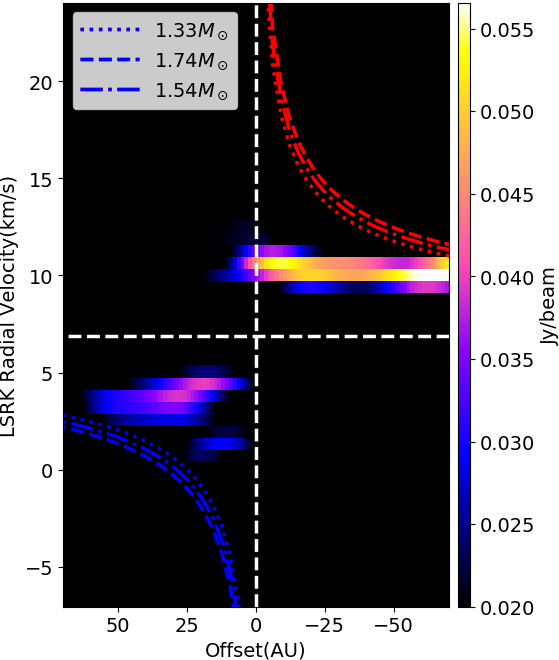}{0.35\textwidth}{b}
    \label{fig:pv}
	}
	\caption{(a) ALMA CO 2-1 moment 0 image of FS Tau A, including both blue and red-shifted component. The red line shows the radius and position angle of the generated P-V map. (b) P-V map of FS Tau A ALMA CO 2-1 image central region. The image is drawn along a slit with a position angle of 9.9$^\circ$, a radius of 0.$\arcsec$5 (70 AU), and a width of 1 pixel. The blue and red lines show the blue- and red-shifted Keplerian rotation curves, respectively, assuming a LSRK radial velocity of 6.9 km/s. The horizontal dashed line shows an LSRK radial velocity of 6.9 km/s and the vertical line corresponds to an offset of 0 AU.}
\label{fig:fstaupv}
\end{figure}

Simulations such as \citet{Facchini2017} suggested that in the circumbinary disk systems, the inner region of the disk could be warped or broken from the outer disk, resulting in a misaligned inner circumbinary disk relative to the outer one. We first investigate that if this can be the case. From Figure~\ref{fig:fstau0comp1}(c) it seems that structure A has relatively large velocity dispersion. To check whether this indicates a near-edge-on protoplanetary disk misaligned with the outer one, we drew a position-velocity (P-V) map of structure A and its opposite red-shifted component structures. In the following discussion, we assume that the inclination of the possible disk is $\sim$90$^\circ$; the conclusions would be similar for other near-edge-on inclinations. The P-V map is drawn along a PA of 9.9$^\circ$, a radius of 0.$\arcsec$5 (70 AU), and a 1-pixel width from the primary star Aa, as shown by the red line in Figure~\ref{fig:fstaupv}(a). From the P-V map, we can see that two different velocity components exist: one is at the bottom left of the P-V map, corresponding to the A structure in the blue-shifted component, and the other is at the top right of the P-V map, corresponding to part of red-shifted component. We also show the Keplerian rotation curve in this image, assuming the central mass range we derived and an LSRK velocity of 6.9 km/s. The rotation map can barely fit the P-V map at the outer radius. However, in the inner region, e.g., at radius $\sim$20 AU, both components keep their velocities at the outer radius and are slower than the Keplerian velocity. 

In the P-V map, no component with a velocity larger than the Keplerian velocity is observed. The velocities of the structures do not fit well with the rotation velocity. In this case, the existence of a disk is doubtful. The outer boundary of structure A is only at 0.$\arcsec$4 or 56 AU, and the semi-major axis of this binary system is about 28-46.9 AU. Therefore, even if the slower velocity in the inner radius is due to the viscosity from the gas, it is a circumbinary disk around both stars instead of a circumstellar disk. For the disk to exist, the gap size opened by the binary should be smaller than the circumbinary disk. 

The gap size relative to the binary orbital semi-major axis depends on the binary mass ratio, binary orbital eccentricity and misaligned angle between the binary orbit and the disk. The orbital inclination range of FS Tau A in our MCMC result is about 32.2$^\circ$-55.5$^\circ$. Therefore, if we assume that the disk is edge-on (e.g., inclination equals to 90$^\circ$) to us, the misaligned angle between the disk and the binary orbital plane is 34.5$^\circ$-57.8$^\circ$ and the orbital eccentricity is about 0.22-0.512. \citet{Miranda2015} suggested that in circumbinary disk system, a binary with misaligned angle of 45$^\circ$ relative to the circumbinary disk -- close to our case should open a gap at least twice of the binary semi-major axis, i.e., larger than 56 AU. Therefore, we suggest that such a misaligned inner circumbinary disk does not exist, i.e., structure A does not indicate a protoplanetary disk.

\subsection{Streamers?}

As mentioned in Section 1, a binary can open gaps in its circumbinary disk. Streamers, which usually come in pairs, can penetrate this gap, sustaining the circumstellar disks around each of the binary stars. After comparing our observations with simulated streamers, such as those reported by \citet{Farris2014} and \citet{Nelson2016}, we found that the shape of the detected structures, especially the near-infrared structures, resembles that of simulated streamers around binaries, which makes us believe that these structures are actually streamers triggered by the binary. We consider that the S1 structure and the blue-shifted component indicate a streamer to the star Aa since structure A in the sub-millimeter band is connected to Aa. The north arm S2 and the red-shifted component indicate another streamer to Ab. Regarding the very bright part shown in the red-shifted component near the binary, we suggest that it represents the streamer connecting the binary.

\begin{figure}[ht!]
	\figurenum{11}
    \gridline{
    \fig{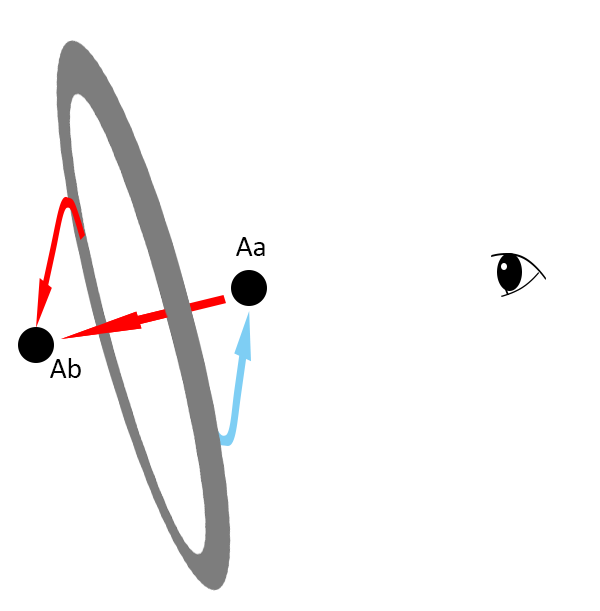}{0.5\textwidth}{a}
    \label{fig:illu1}
    \fig{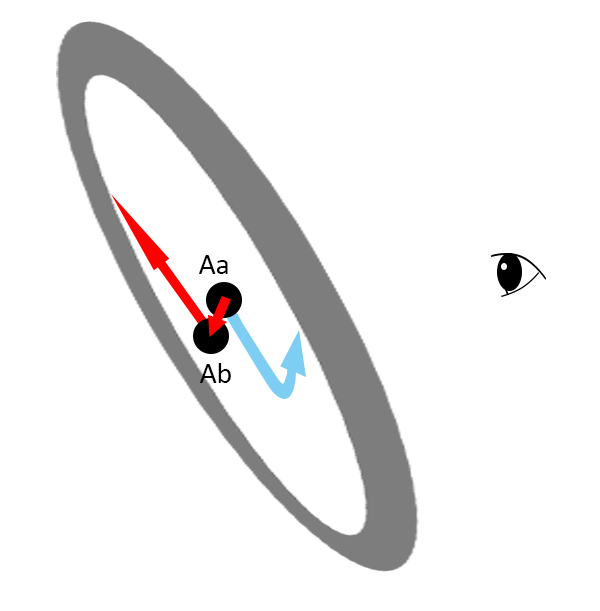}{0.5\textwidth}{b}
    \label{fig:illu2}
	}
	\caption{Two suggested interpretations for the structures around FS Tau A. (a) Misaligned streamer hypothesis. The binary has a relatively large misalignment angle to the circumbinary disk. FS Tau Aa is located in front of the bottom right side (the southeast side in the observed structures) of the circumbinary disk so that observers on Earth (shown as an aye) see a blue-shifted streamer to Aa from the right. FS Tau Ab is located behind the top left side (northwest side in the observed structures) of the circumbinary disk so that observers on Earth see a red-shifted streamer to Ab. The streamer between Aa and Ab is expected to be from Aa to Ab since the connection between the stars is red-shifted. (b) Outward-moving streamer hypothesis. The binary has a small misalignment angle to the circumbinary disk as previous calibrations suggested, but the materials in the streamers are moving outwards to the disk due to the torque caused by the binary, so observers on Earth will see that the structures in the southeast are blue-shifted and those in the northwest are red-shifted.}
    \label{fig:illu}
\end{figure}

Based on ALMA observations, the southeast structure shows blue-shifted characteristics, indicating that the materials in S1 are moving toward Earth. The red-shifted component indicates that some materials in the north are moving away from Earth. In Figure~\ref{fig:radialmap}, the southeast structure S1 is brighter than the north structure S2, which is consistent with the image derived by \citet{Hioki2011}, in which the southeast part of the disk is brighter than the northwest part. Considering that the disk is large ($\sim$600 AU), we suggest that this difference in brightness is simply due to the southeast side being nearer to Earth, rather than that the dust surface density in the southeast part being much higher. If we assume that the streamers are flowing to the stars, then star Aa should be in front of the southeast side of the outer circumbinary disk, so that the streamer will move toward Earth to reach Aa, and Ab should be farther than the northwest side of the disk, so that the materials will move away from Earth to reach it. In a simple model of this situation, the outer circumbinary disk is highly misaligned with the binary orbit, Aa is in front of the disk, and Ab is behind the disk, as shown in Figure~\ref{fig:illu}(a). Simulations such as that by \citet{Nixon2013} have proven that streamers to a binary can happen in a disk-binary misaligned system with a misalignment angle of up to 60$^\circ$, supporting this scenario.

To make this scenario true, the southeast inner edge of the circumbinary disk should be behind Aa while the northwest inner edge of the circumbinary disk should be in front of Ab. \citet{Hioki2011} suggested that the $\sim$600 AU circumbinary disk around FS Tau A has a PA of 15$^\circ$-40$^\circ$, an inclination of 30$^\circ$-40$^\circ$, and its southeast side nearer to Earth. While the binary has a semi-major axis 28-46.9 AU, inclination 32.2$^\circ$-55.5$^\circ$. In this case, if we consider the circumbinary disk is ``flat", i.e., the disk is not severely truncated by the binary, the southeast part of the circumbinary disk, even at 140 AU, is closer to Earth than the binary, so this configuration conflicts with this scenario. However if the inner region of the circumbinary disk is severely truncated by the binary, it is still possible that the streamers are moving to the stars. 

Another interpretation is that the streamers are moving away from the stars. A simulation by \citet{Shi2012} suggests that while some materials move toward the central binary black hole, some other materials in the streamers move out from the stars to the disks; this should also happen in circumbinary disk systems. Considering FS Tau A, the southeast side of the disk is nearer to Earth, so observers on Earth will see the materials moving from Aa show a blue shift and those from Ab show a red shift, as shown in Figure~\ref{fig:illu}(b). This interpretation fits well with the current suggested inclinations of binaries and disks given by \citet{Tamazian2002} and \citet{Hioki2011}. We thus consider this to be the best explanation of the detected structures now. However, it does not indicate that materials will keep moving from the stars and there will be no circumstellar disks around the each of the binary at all. Simulations such as \citet{Munoz2016} show that the streamers triggered by the binary vary with time, even in one period, streamers are destroyed and regenerated, while material may be moving towards the star at some time, and moving away from the star at some other time. Therefore, it is difficult to connect the instantaneous kinematics in the streamers to the observed accretion rate onto the star, as the two can have a phase difference. There could still be a compact circumstellar disk around the star to sustain accretion, replenished by the streamers and other cross-cavity flows from time to time, but planet formation in such hostile environment might be hard.

Gravitational instability may also produce large-scale spirals and streamers in disks with sufficiently high mass ($M_{disk}/M_{star}>\sim0.1$; e.g., \citet{Kratter2016,Vorobyov2005,Vorobyov2010}). These structures can be visible in near-infrared scattered light and millimeter dust emission \citep{Dong2016b}. If the disk can cool efficiently, it may also fragment to form self-gravitating objects, including protostars \citep{Gammie2001,Rafikov2009}. The protostar system L1448 IRS3B may be an example of this \citep{Tobin2016}. The Subaru Telescope image of the FS Tau A system appears to have a similar morphology. However, the total millimeter flux of the system (2.72 mJy at 1.3 mm) indicates a low disk mass, $M_{disk}/M_{star}\sim0.5$ \citep{Akeson2019}, assuming a dust opacity of $\tau_{1.3mm} = 2.3 cm^2/g$ \citep{Beckwith1990}, an average dust temperature of 20 K, and a gas-to-dust mass ratio of 100:1. The measured accretion rate is more than an order of magnitude smaller than the expected value for a gravitationally unstable disk ($\dot{M}>\sim10^{-7}M_\odot/ yr$; \citet{Dong2015}). We thus consider that the observed structures in the FS Tau A system are unlikely to have been produced by gravitational instability.

\section{Conclusion}

Our near-infrared polarimetric observations as well as archived CO 2-1 emission data resolved the disk structures near the young binary FS Tau A. An analysis of the data revealed the following: 

1. In the near-infrared H-band, we found one arm-like structure (S1) at the southeast side of the binary, which starts at the south of the binary and then turns to the northeast, extending to at least $\sim$1.$\arcsec$5 or 210 AU from the binary. We found another structure (S2) extending to at least $\sim$1.$\arcsec$7 or 238 AU north of the binary. 

2. At sub-millimeter wavelengths, we found two structures within an LSRK velocity range of 1.19-5.64 km/s: structure A extends from Aa to about 56 AU in the south, and structure B is located at about 21 AU from Ab in the southeast and has length about 98 AU. For the LSRK velocity range of 8.17-16.43 km/s, one structure was discovered; it extends to $\sim0.\arcsec7$ (98 AU) north from the binary, and then turns to $\sim0.\arcsec6$ (84 AU) west. 

3. According to the morphology and velocity of the detected structures, we suggest that these structures are most likely two streamers connecting the binary and its circumbinary disk. Current calibrated inclinations of the binary and the circumbinary disk suggest that the materials in the streamers are moving from the binary to the surrounding disk. This discovery is helpful for understanding the dynamics of the streamers in the circumbinary disk system, and reveal the formation and evolution of the surrounding disk around each star in the binary system.

\acknowledgments
We are grateful to an anonymous referee for constructive suggestions that improved our paper. This paper used data collected by the Subaru Telescope, which is
operated by the National Astronomical Observatory of Japan (NAOJ) and the National Institutes of Natural Sciences (NINS). 
We thank the Subaru Telescope staff for their support during the observations. This paper made use of the following ALMA data: ADS/JAO.ALMA\#2013.1.00105.S. ALMA is a partnership of ESO (representing its member states), NSF (USA), and NINS (Japan), together with NRC (Canada), NSC, ASIAA (Taiwan), and KASI (Republic of Korea), in cooperation with the Republic of Chile. The Joint ALMA Observatory is operated by ESO, AUI/NRAO, and NAOJ. We would also like to acknowledge the access given to us to the SIMBAD database operated by the Strasbourg Astronomical Data Center (CDS), Strasbourg, France. 
IRAF is distributed by the National Optical Astronomy Observatory, which is operated by the Association of Universities for Research in Astronomy (AURA) under a cooperative agreement with the National Science Foundation. M.T. is supported by a Grant-in-Aid for Scientific Research (no. 15H02063).  E. A. is supported by MEXT/JSPS KAKENHI grant no. 17K05399.

\bibliographystyle{yahapj}
\bibliography{references}

\end{document}